# ChatGPT in education: A discourse analysis of worries and concerns on social media

Lingyao Li[1]*, Zihui Ma[2], Lizhou Fan[1], Sanggyu Lee[2], Huizi Yu[1], Libby Hemphill[1]
[1]School of Information, University of Michigan, Ann Arbor, MI
[2]Department of Civil and Environmental Engineering, University of Maryland, College Park, MD
*Corresponding author, lingyaol@umich.edu

**Abstract**

The rapid advancements in generative AI models present new opportunities in the education sector. However, it is imperative to acknowledge and address the potential risks and concerns that may arise with their use. We analyzed Twitter data to identify key concerns related to the use of ChatGPT in education. We employed BERT-based topic modeling to conduct a discourse analysis and social network analysis to identify influential users in the conversation. While Twitter users generally expressed a positive attitude towards the use of ChatGPT, their concerns converged to five specific categories: academic integrity, impact on learning outcomes and skill development, limitation of capabilities, policy and social concerns, and workforce challenges. We also found that users from the tech, education, and media fields were often implicated in the conversation, while education and tech individual users led the discussion of concerns. Based on these findings, the study provides several implications for policymakers, tech companies and individuals, educators, and media agencies. In summary, our study underscores the importance of responsible and ethical use of AI in education and highlights the need for collaboration among stakeholders to regulate AI policy.

**Keywords**

ChatGPT; Education; Social media; Topic modeling; Sentiment analysis; Social network.

# 1. Introduction

The development of large language models (LLMs) has marked significant progress in the domain of generative artificial intelligence (AI) (Fan et al., 2023; Kasneci et al., 2023). ChatGPT, a chatbot that was launched in November 2022 (OpenAI, 2022b), is one such generative AI model that has shown tremendous potential in language understanding and knowledge retention. Multiple evaluations and tests have validated its capabilities (Dwivedi et al., 2023; Li et al., 2023), particularly in the realm of higher education (Fauzi et al., 2023; Firat, 2023). For example, ChatGPT was able to pass graduate-level exams from law and business schools (Kelly, 2023), and the recently released GPT-4 model achieved top 10% in a law test (Koetsier, 2023). ChatGPT's performance has even been recognized in academic publications, with several papers listing it as a co-author (Mintz, 2023).

ChatGPT has brought about opportunities for education. Its ability to provide personalized learning experiences, to assist in creating educational content, and to overcome language barriers can have a significant impact on teaching and learning outcomes (Adiguzel et al., 2023; Chen, 2023). For example, ChatGPT can help teachers generate questions, quizzes, assignments, and interactive educational content, such as games and simulations, that cater to the students' learning styles (Kasneci et al., 2023; Lee, 2023). ChatGPT can also support students to customize learning and provide feedback accordingly (Kasneci et al., 2023).

However, the use of ChatGPT in education has raised potential concerns and risks (AlAfnan et al., 2023; Kasneci et al., 2023; Sok & Heng, 2023). One concern is the ethical implications of ChatGPT's ability to write scientific essays (Mhlanga, 2023), which may compromise the authenticity and originality of research (Malik et al., 2023). Another issue is the use ChatGPT by students to outsource their writing (Lund et al., 2023), which poses a challenge for academic institutions that rely on plagiarism detection tools to maintain academic integrity (Fijačko et al., 2023; Ventayen, 2023) and potentially undermines students' writing skill development (Kasneci et al., 2023; Sallam, 2023). In addition, the output of ChatGPT can be biased or nonsensical, possibly leading to the dissemination of incorrect information (Baidoo-Anu & Owusu Ansah, 2023; Choi et al., 2023).

The implementation of ChatGPT in education has sparked a large-scale conversation on social media (Kelly, 2022; Taecharungroj, 2023), allowing individuals to exchange information and accelerate knowledge dissemination. Social media's ability to quickly and broadly disseminate information can facilitate the emergence of critical opinions in a short period, which can be valuable for decision-makers to address concerns (Haque et al., 2022; Li et al., 2022). In addition, social media platforms can promote knowledge sharing and collaboration among policymakers, tech companies, educators, and students (Ahmed et al., 2019), facilitating the development of best practices for responsible AI in education. For example, educators can share their experiences and concerns about integrating AI into the learning process, while tech companies and engineers can offer insights into the latest developments in AI models and strategies.

Analyzing crowdsourced opinions through social media offers two significant advantages. First, compared to expert opinions, crowdsourcing provides a more comprehensive and diverse perspective on how the general public perceives concerns based on their experiences or observations. Second, dissecting the social network on social media can reveal important users that are frequently implicated in the conversation. Therefore, this study proposed the following two research questions.

- **RQ1 (Concerns)**: What are the key concerns that Twitter users perceive with using ChatGPT in education?
- **RQ2 (Accounts)**: Which accounts are implicated in the discussion of these concerns?

To address the research questions, we conducted a discourse analysis of Twitter data pertaining to the use of ChatGPT in education. Leveraging topic modeling, we were able to cluster negative sentiment tweets and identify the concerns perceived by Twitter users. Using social network analysis, we were able to investigate opinion leaders and frequently implicated users in the conversation. Through this analysis, our study aims to inform stakeholders of the prevailing concerns associated with the use of ChatGPT in education and to highlight their responsibilities in mitigating potential risks. Our study further emphasizes the crucial role of collaboration among stakeholders towards the development of effective strategies to ensure the responsible and ethical use of generative AI in educational settings.

## 2. Literature review

### 2.1. The use of Generative AI models in education

Generative AI models have garnered significant interest and attention from the public due to their ability to produce content that closely resembles human-generated content. These models can respond to complex and diverse prompts, including images, text, and speech (Dwivedi et al., 2023). Among them, ChatGPT (OpenAI, 2022b) and DALL-E (OpenAI, 2022a) are the two popular GPT-based AI products released by OpenAI in 2022. Other Generative AI models like Stable Diffusion from Stability.ai and Lensa have the ability to create user portraits known as Magic Avatars (Pavlik, 2023). Google has recently released a new Generative AI system called Bard powered by Language Model for Dialogue Applications (LaMDA) (Pichai, 2023).

These Generative AI models have emerged as promising tools for enhancing learning and teaching processes in education (Dwivedi et al., 2023). A recent study demonstrated the potential of GPT-3 to generate multiple-choice questions and answers for reading comprehension tasks (Dijkstra et al., 2022). Similarly, Bhat et al. (2022) proposed a pipeline for generating assessment questions based on a fine-tuned GPT-3 model to facilitate self-learning. Moreover, conversational agents such as Blender and GPT-3 have been explored for educational dialogues, generating conversational dialogues that convey a deep understanding of the learner (Tack & Piech, 2022). Beyond above applications, Generative AI models have been demonstrated useful in various educational tasks, such as generating code explanations (MacNeil et al., 2022), writing essays (Park et al., 2022), and providing formative feedback on student work (Jia et al., 2021).

### 2.2. Opportunities and risks of using generative AI models in education

Generative AI could benefit the educational sector with personalized and innovative teaching and learning methods. Prior studies have shown that these models can help educators create teaching materials such as quizzes, tests, and worksheets (Abdelghani et al., 2022; Dijkstra et al., 2022; Gabajiwala et al., 2022). They can also help analyze student data, identify learning patterns, and generate valuable insights to refine teaching methods (Bernius et al., 2022; Moore et al., 2022; Zhu et al., 2020). For students, generative AI models can provide interactive and engaging learning experiences that cater to their learning styles and needs. One typical application is to summarize information from multiple sources, which can aid the process of knowledge acquisition and improve learning efficiency (Haleem et al., 2022). They can also benefit students with disabilities by enhancing accessibility and developing more inclusive learning strategies. For example, Kasneci et al. (2023) discussed how ChatGPT can be used with text-to-speech or speech-to-text technologies to assist students with hearing or visual impairments.

Although their application is on the rise, concerns have emerged about generative AI's impacts on students, educators, and the educational landscape. Researchers have highlighted the potential risks associated with students' reliance on AI-generated content, which may hinder their critical thinking and problem-solving skills (Iskender, 2023; Kasneci et al., 2023). Additionally, the authenticity and originality of AI-generated content have been questioned, as generative AI models can produce content that mimics human writing, raising concerns about academic integrity (Alser & Waisberg, 2023; Kasneci et al., 2023; Lim et al., 2023). Educators may also be hesitant to adopt generative AI in their teaching practices due to concerns about job security, lack of understanding of the technology, or fear of its potential negative impacts on the education system (Atlas, 2023). To better understand these concerns, we compiled a list of prior work that discusses the ethical and practical concerns of using ChatGPT in education (see **Table 1**).

**Table 1**. Existing concerns regarding the implementation of ChatGPT in education.

| Author and Year | Concerns |
| --- | --- |
| AlAfnan et al. (2023) | Discourage writing, impact on student learning and development, challenge to evaluate learning outcomes. |
| Atlas (2023) | Risk of plagiarism, proper attribution (e.g., the source of information), workforce displacement and reskilling; multiple concerns and suggestions for educators (e.g., academic integrity, data privacy and security, ethical considerations, accessibility, transparency, institutional policies, professional development). |
| Baidoo-Anu and Owusu Ansah (2023) | Lack of human interaction, limited understanding, bias in training data, lack of creativity, lack of contextual understanding, limited ability to personalized instruction, privacy. |
| Choi et al. (2023) | Failure to generate sufficient detail, misunderstanding of terms, departure from the material. |
| Halaweh (2023) | Concerns from text generation (e.g., writing, editing), concerns from ideas generation (e.g., critical thinking, originality). |
| Kasneci et al. (2023) | Copyright issues, bias and fairness, impact on critical thinking and problem-solving, lack of understanding and expertise, difficulty to distinguish machine- and human-generated text, cost of training and maintenance, data privacy and security. |
| Megahed et al. (2023) | The quality of AI outputs, biased prediction, ethical questions. |
| Mhlanga (2023) | Respect for privacy, transparency, responsible AI (AI limitations), accuracy of information, replacement of teachers, fairness and non-discrimination. |
| Qadir (2022) | Incorrect information and mistakes, unethical conduct, potential to exacerbate inequalities. |
| Rudolph et al. (2023) | Threat to essay, the relevance or accuracy of the generated information, replacement of teaching jobs. |
| Sallam (2023) | Ethical, copyright, transparency, and legal issues, the risk of bias, plagiarism, lack of originality, inaccurate content with risk of hallucination, limited knowledge, incorrect citations, cybersecurity issues, and risk of infodemics. |
| Sallam et al. (2023) | Risk of plagiarism, copyright issues, the risk of academic dishonesty, lack of personal and emotional interactions, suppressing the development of critical thinking and communication skills. |
| Sok & Heng (2023) | Academic integrity, unfair learning assessment, inaccurate information, over-reliance on AI. |
| Thorp (2023) | Scientific misconduct. |

| Tlili et al. (2023) | Ethical concerns, response quality, creation of misinformation, data security risk. |
| Thurzo et al. (2023) | Impact on creativity and skills, cheating and plagiarism. |
| Zhai (2022) | Biased outcomes, privacy, replacement of human jobs, lack of transparency. |

## 2.3. Social media discussion of ChatGPT

Most of the studies we reviewed rely on authors' observations or a review of existing literature to identify concerns, which may limit the diversity of perspectives and lack real-world relevance of the discussion. Social media offers a platform for users to share their opinions based on first-hand observation or experience, allowing us to tap into current trends and issues related to the use of ChatGPT and gain a more comprehensive collection of the concerns.

A few attempts have leveraged social media data to crowdsource opinions regarding the application of ChatGPT (Feng et al., 2023; Haque et al., 2022). For example, Leiter et al. (2023) collected and analyzed Twitter data to investigate people's perceptions of ChatGPT. They found that ChatGPT has been generally well-received on social media, particularly in scientific domains like medical fields, but has been viewed as a threat to the education sector. In another study, Taecharungroj (2023) used Latent Dirichlet Allocation (LDA) topic modeling to examine tweets following ChatGPT's launch. The study identified several domains in which ChatGPT could operate effectively, including creative writing, essay writing, prompt writing, code writing, and question answering. However, the study also highlighted the potential impacts of ChatGPT on both technology and humans, such as concerns about job displacement.

Two other studies leveraged discussions from Twitter and TikTok about ChatGPT in education. Tlili et al. (2023) analyzed tweets about ChatGPT using social network analysis and conducted interviews to understand stakeholders' perceptions and users' experiences. Their findings revealed several concerns of using ChatGPT, including cheating, honesty, privacy, and manipulation. Haensch et al. (2023) collected TikTok videos to investigate students' usage and perceptions of ChatGPT. They found that TikTok videos promoted ChatGPT for academic assignments and discussed ways to deceive AI detectors. However, they noted a lack of videos depicting ChatGPT-generated nonsensical or unfaithful content. The authors suggested that this gap in information could affect educators' attitudes towards using ChatGPT for teaching and grading.

Our review of existing studies has revealed several research gaps. As mentioned, many studies have relied on expert opinions or literature reviews to summarize concerns related to this topic. This approach may not capture diverse perspectives or reflect a more general discussion from the public. While a few studies have leveraged social media data to crowdsource opinions, two questions remained unclear, as listed in the **Introduction**. Our study stands apart from the two relevant social media studies in two ways. First, we applied Natural Language Processing tools to mine opinions from a vast collection of tweets, thereby providing insights from a broad cross-section of Twitter users. Second, we analyzed influential users and their profile information from the social network, further providing practical implications.

To address these gaps, we first used topic modeling to identify concerns expressed in negative-sentiment tweets to gain insights into the public's concerns about using ChatGPT in educational settings. Second, we employed social network analysis to identify opinion leaders and frequently mentioned accounts in the conversation. This information can provide useful insights into who the crowd identifies as stakeholders for the application of ChatGPT in education.

## 3. Data and methods

**Figure 1** displays the graphical illustration of our research framework. The process started with data collection, which involved collecting and archiving tweets written in English related to the subject of ChatGPT in education, as elaborated in **Section 3.1**. Subsequently, we used a sentiment model developed on the RoBERTa architecture to classify the sentiment, as detailed in **Section 3.2**. Thereafter, we employed the BERTopic tool to cluster negative tweets into distinct topics, with the aim of eliciting potential concerns from negative tweets, as described in **Section 3.3**. Next, we used social network analysis to chart the dissemination network among users, which helps identify key accounts that propagate the concerns and should be apprised of the associated risks, as described in **Section 3.4**.

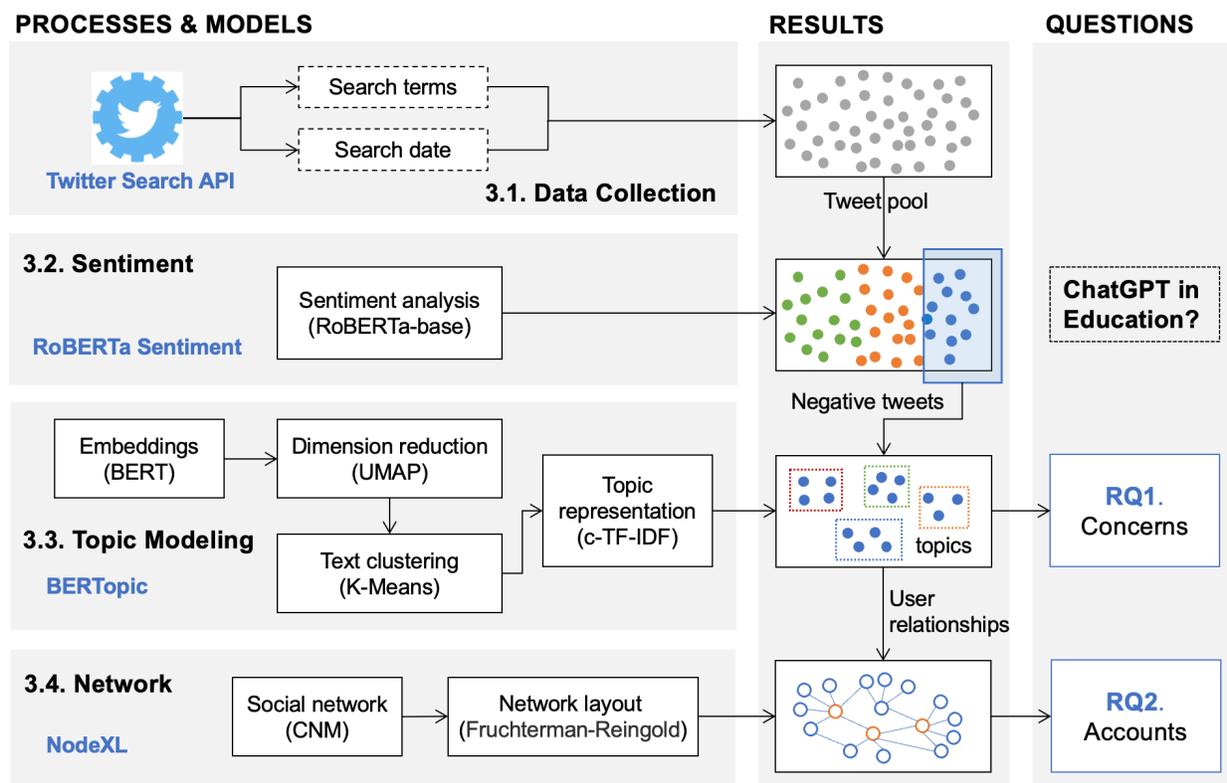

**Figure 1**. Illustration of the research framework.

### 3.1. Data collection

We used the academic version of Twitter Search API and selected fourteen education-related search terms along with the keyword "ChatGPT" to collect relevant tweets. Our search covered tweets posted from December 1, 2022, to March 31, 2023. OpenAI released ChatGPT on November 30, 2022 (OpenAI, 2022b). Specific search parameters are provided in **Table 2**. To ensure consistency in our dataset, we limited our search to English tweets that contained the designated keywords. This decision was based on the capabilities of the NLP tools we used to analyze the tweets. As a result, we collected a total of 247,484 tweets, including original tweets, mentions, replies, and retweets, in which 84,828 were original tweets.

Table 2. Twitter API search conditions.

| Conditions | Search |
|---|---|
| API tool | Twitter Search API for academic research |
| Search date | December 1, 2022 to March 31, 2023 |
| Search terms | ChatGPT + school; ChatGPT + college; ChatGPT + university; ChatGPT + education; ChatGPT + student; ChatGPT + teacher; ChatGPT + learning; ChatGPT + curriculum; ChatGPT + class; ChatGPT + exam; ChatGPT + homework; ChatGPT + teaching; ChatGPT + academia; ChatGPT + academic. |

## 3.2. Sentiment analysis

For sentiment analysis, we applied the "twitter-roberta-base-sentiment-latest" released by Loureiro et al. (2022), which is a built based on its predecessor "twitter-roberta-base-sentiment" model released by Barbieri et al. (2020). Barbieri et al. (2020) selected RoBERTa (Liu et al., 2019) as a pre-training approach due to its top performance in the General Language Understanding Evaluation benchmark (GLUE). RoBERTa is a robustly optimized Bidirectional Encoder Representations (BERT) pre-training approach. It is particularly suitable for Twitter where most tweets are composed of a single sentence (Devlin et al., 2019). In addition, compared to context-free embedding models such as Word2Vec and GloVE, BERT embedding is based on the transformers architecture and relies on an attention mechanism to generate contextualized representation based on the text.

The RoBERTa-base sentiment model was fine-tuned for sentiment analysis with the TweetEval benchmark, which specifically focuses on analyzing tweet sentiment. Barbieri et al. (2020) added a dense layer to reduce the dimensions of RoBERTa's last layer to the number of labels in the classification task to prepare the model for sentiment classification. This model has demonstrated its superior performance over FastText and Support Vector Machine (SVM) -based models with n-gram features (Barbieri et al., 2020). Loureiro et al. (2022) further updated the model by training it on a larger corpus of tweets, based on 123.86 million tweets extracted until the end of 2021, compared to its predecessor's 58 million tweets.

By leveraging the latest model, we were able to analyze the sentiment of tweets. Tweet examples and their resulting sentiment classifications are presented in **Table 3**. The sentiment classification follows the highest score returned by the softmax layer in the RoBERTa model. We used only the tweets with negative sentiment (N = 70,318) in our analyses.

Table 3. Examples of tweet sentiment classification.

| Tweet | Sentiment Score | Sentiment |
|---|---|---|
| It's easy to underestimate the impact on education ChatGPT will have. I asked my kids what concept they were struggling with understanding at school. Like we've seen with YouTube or Khan Academy, supplementing their edu with tools like this can make them smarter than we ever were. | Negative: 0.094 Neutral: 0.536 Positive: 0.370 | Neutral |
| #ChatGPT is fascinating for AI in our conversations with customer support, assistants etc. to write our homework, blog articles, grant proposals etc. Some even speculate that we may be a | Negative: 0.010 Neutral: 0.229 Positive: 0.761 | Positive |
| few years away from this technology replacing search engines like #Google. | | |
| I once had a great engineering professor (Hi Prof. Lasky!). But he was old school. He introduced himself with this photo, and would assign problems that we were required to solve with a slide rule.<br><br>This is a thread about #ChatGPT and AI tool adoption in math/engineering. | Negative: 0.016<br>Neutral: 0.211<br>Positive: 0.773 | Positive |
| The software recommendations that I have seen so far from chatgpt are often incorrect.<br><br>Sometimes obvious (an incorrect method name) but also sly (timing). These mistakes are often caught by tools but people say this is a tool.<br><br>In reality: it's homework. Not buying today. | Negative: 0.826<br>Neutral: 0.166<br>Positive: 0.008 | Negative |
| Professors Caught Students Cheating on College Essays With ChatGPT. | Negative: 0.762<br>Neutral: 0.228<br>Positive: 0.010 | Negative |

## 3.3. BERT-based topic modeling

Topic modeling enables the identification of semantic themes in vast quantities of text data, such as social media data. We employed the BERTopic tool, which involves applying BERT embedding to extract semantically relevant sentence embeddings from tweets. As mentioned, BERT embedding offers a distinct advantage due to its contextualized representation. Specifically, we utilized the Sentence-BERT (SBERT) Python package for this task (Reimers & Gurevych, 2019).

Given that the BERT embedding converts tweets into high-dimensional vectors, we employed a dimensionality reduction technique called Uniform Manifold Approximation and Projection (UMAP), as proposed by McInnes et al. (2018). UMAP can help mitigate the "curse of dimensionality" while preserving the local and global structure of the dataset. This feature of UMAP is particularly useful for constructing topic models that rely on the structural similarities of word vectors (McInnes et al., 2018).

After that, we employed the Scikit-learn Python package to apply K-Means clustering and group similar sentence embeddings into topics (Buitinck et al., 2013). We chose K-Means clustering because of its simplicity, computational efficiency, and effectiveness in handling large datasets. We experimented with various cluster numbers, including 50, 100, and 200, to identify the optimal number of clusters for our analysis. Our results revealed that 50 or 100 clusters produced clusters that mixed several topics. 200 clusters provided more coherent and separable clusters.

The final stage of our topic modeling process involves the representation of topics. We used the count vectorizer within the Scikit-learn Python package to tokenize the topics. We then applied class-based Term Frequency-Inverse Document Frequency (c-TF-IDF) to extract the topical keywords and representative tweets from each cluster (Grootendorst, 2022). This process enabled us to interpret the themes from clustered topics and assign different topics into a theme (i.e., "category" in our following analysis).

## 3.4. Social network analysis

**Figure 2(a)** illustrates the relationships among Twitter users based on mentions, retweets, quotes, and replies. We focused on two types of interactions: mentions and retweets. A mention is a tweet that includes another user's name in the text, and when a user is mentioned, they receive a notification from Twitter. A retweet is a repost of another user's tweet (Twitter Inc, 2023). By analyzing mentions, we aimed to identify users who were frequently implicated in the conversation and thus should be responsive to the concerns. By analyzing retweets, we aimed to identify the opinion leaders in communicating risks and concerns to a broader audience.

We utilized degree centrality to investigate communication patterns in Twitter social networks (Powell, 2015). **Figure 2(b)** illustrates the degree of centrality of each Twitter user, which reflects the number of connections (edges) a user has with other users (vertices). Graph theory introduces three types of centralities in a network graph: in-degree, out-degree, and betweenness centrality (Powell, 2015). We focused on in-degree centrality in this study as it allows us to identify the influential users (those who receive attention from other users).

To analyze the social network of Twitter users, we employed NodeXL, a network analysis and visualization software package embedded with Microsoft Excel (Smith et al., 2010). With NodeXL, we selected the Clauset, Newman, and Moore (CNM) algorithm to plot the community structure (Clauset et al., 2004) and the Fruchterman-Reingold layout to display the network (Fruchterman & Reingold, 1991).

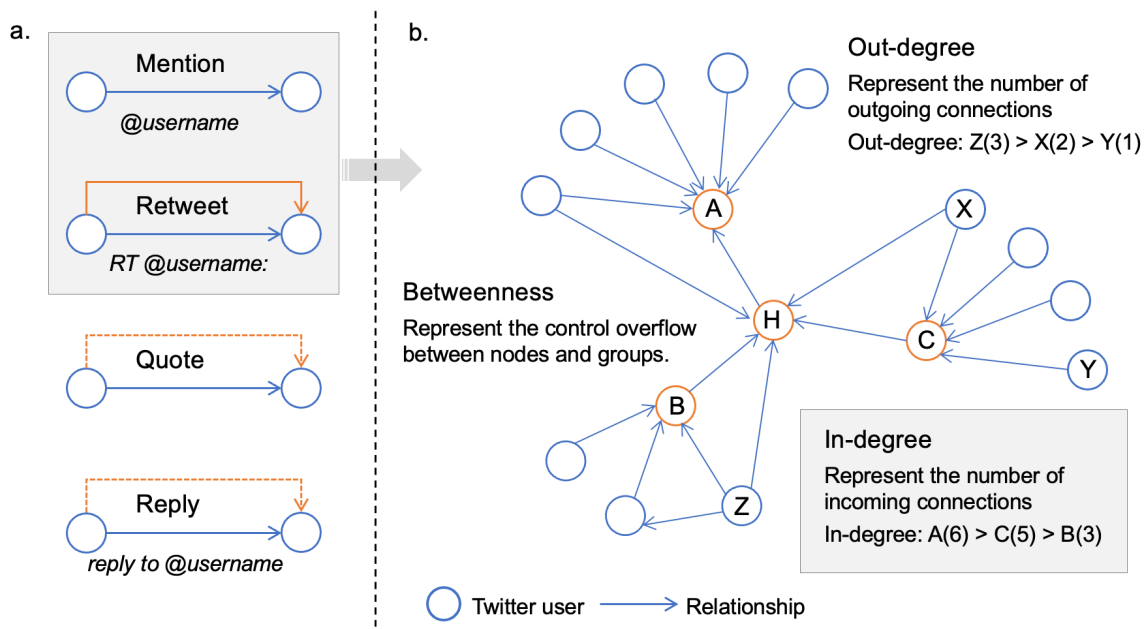

**Figure 2**. Illustrations of Twitter interactions and degree centrality in Graph Theory (adapted from Kim et al. (2018) and Li et al. (2021)). (a) Types of Twitter interactions. (b) Degree centrality.

## 4. Results

The results section consists of three parts. In **Section 4.1**, we used the RoBERTa tool to analyze sentiment and then identify events that were associated with sentiment changes. This sets the context for the discourse analysis and provides an overview of sentiment trends. In **Section 4.2**, we utilized BERTopic modeling to perform discourse analysis. This section addressed **RQ1** by identifying the concerns that garnered the most attention within Twitter communities. In **Section 4.3**, we employed the NodeXL tool to describe the social network of Twitter users who engaged in the conversation. This section answers **RQ2** by identifying influential users.

### 4.1. Sentiment trends and analysis

**Figure 3** displays the sentiment trend during the study period. Out of the 247,484 tweets, 49,528, 127,638, and 70,318 tweets were identified as negative, neutral, and positive, respectively. Similarly, out of the 84,828 original tweets, 16,011 were negative, 42,495 were neutral, and 26,322 were positive. The sentiment analysis suggests that Twitter users have a generally positive attitude towards the application of ChatGPT in education, as indicated by the orange line consistently positioned above the blue line in **Figure 3**.

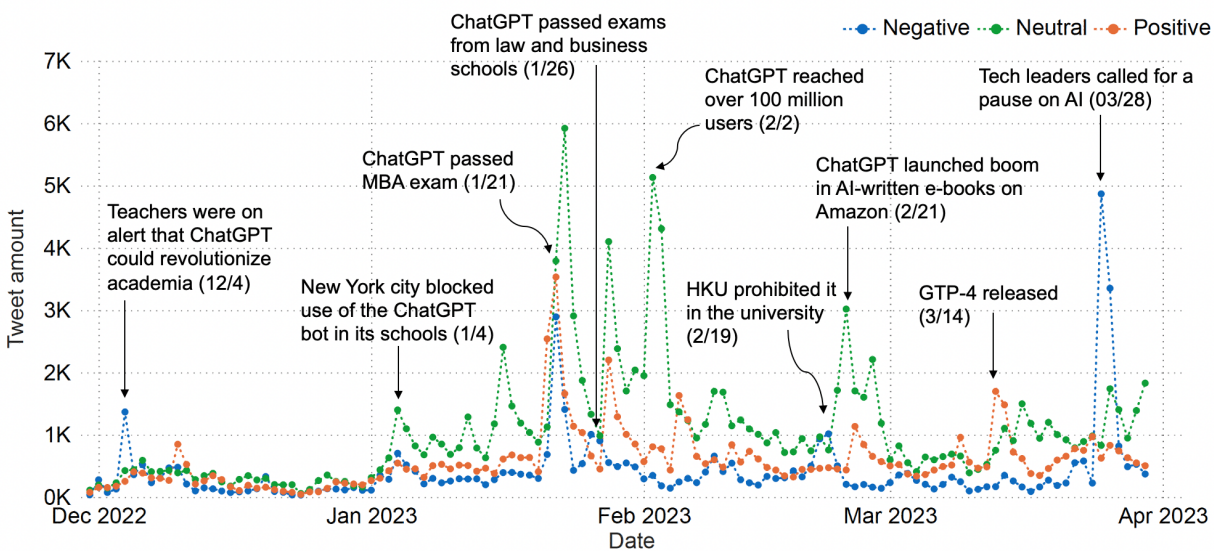

**Figure 3**. Twitter sentiment trend and significant events during the study period.

To better understand the sentiment trends, we identified and annotated a few noteworthy events that likely triggered widespread discussion on Twitter. One such event occurred on March 14, 2023, with the release of OpenAI's highly anticipated GPT-4 model (OpenAI, 2023), resulting in a positive sentiment across Twitter. Similarly, when ChatGPT exceeded 100 million users in early February (Milmo, 2023), it sparked a wide discussion that persisted for several days on Twitter.

However, we also observed several significant events associated with negative sentiment in public attitudes. The first one occurred around December 4, 2022, at the beginning of the study period, as many tweets discussed the potential negative effects of ChatGPT on the education system (Meckler & Verma, 2022; Stokel-Walker, 2022). Teachers expressed concern that this technology could change academia, leading to cheating and other negative impacts on learning. Later on, in late January, these worries resurfaced as news broke that ChatGPT had passed MBA exams and other tests from law and business schools (Kelly, 2023; Rosenblatt, 2023). Furthermore, schools and universities announcing ChatGPT bans in January (Johnson, 2023) also contributed to the negative sentiment.

The largest negative sentiment peak was observed at the end of March, specifically on March 28, 2023. Many tech leaders, including Elon Musk and Steve Wozniak, signed an open letter by the Future of Life Institute (Future of Life Institute, 2023), calling for an immediate pause on giant AI experiments like ChatGPT, citing "profound risks to society and humanity" (Hurst, 2023). This event contributed to the dramatic increase in negative sentiment towards ChatGPT. Overall, these events serve as useful markers in contextualizing and interpreting our findings regarding public sentiment towards the use of ChatGPT in education.

## 4.2. Discourse analysis of concerns

We employed BERTopic to cluster 200 topics from 16,011 original negative-sentiment tweets. Subsequently, we manually examined the topics and categorized them into distinct categories. For consistency, we used the term "topic" to refer to the clusters returned by BERTopic and the term "category" to refer to the manually identified category information. Our categorization was informed by two sources. First, we extracted relevant categories from our literature review, as presented in **Table 1** in **Section 2.2**. Drawing on the insights from prior works, we identified six overarching categories that encapsulated the concerns expressed in the negative tweets.

- **(A) Academic integrity**: topics that describe ethical and moral concerns in academic activities, such as unethical practices, scientific misconduct, the risk of academic dishonesty, and various forms of cheating and plagiarism in **Table 1**.
- **(I) Impact on learning outcomes and skill development**: topics that describe negative impact on learning and skill development. This includes the impact on critical thinking, creativity, problem-solving abilities, as well as writing and coding skills in **Table 1**.
- **(L) Limitation of AI capabilities**: topics that describe the limited capabilities of AI-generated information, such as biased outcomes, lack of contextual understanding, limited ability to personalized instruction, inaccurate information, misinformation, misunderstanding, failure to generate sufficient detail in **Table 1**.
- **(S) Security and privacy**: topics that describe data security concerns, such as data privacy and cybersecurity issues in **Table 1**.
- **(W) Workforce challenges**: topics that describe potential impacts on jobs, such as replacement of teachers, devaluation of job training, as well as workforce displacement and reskilling in **Table 1**.

Second, after reviewing the keywords and a sample of tweets from each of the 200 topics, we identified three additional categories, as presented below. We noted that the "(M) Miscellaneous" category includes topics that contain the key search terms but are not relevant to ChatGPT application in education, such as general AI concerns, tech company competition, challenges of deep learning models, or tweets with sentiment classification errors.

- **(G) General negativity**: topics that describe a generally negative attitude towards the application of ChatGPT in education, such as disruption to the education system and exposing weakness of current education.

- **(O) Operation and management issues**: topics that pertain to the operation of ChatGPT, such as the system breakdowns and charges incurred during usage.
- **(P) Policy and social concerns**: topics that describe the policies and social concerns, such as blocks and bans on the use of ChatGPT, as well as sensitive issues that may cause a social impact.
- **(M) Miscellaneous**: These are topics that do not fit into any of the above categories. They may cover general concerns related to AI, deep learning models, data bias, or topics that arise from incorrect sentiment classification.

After finalizing the categories, we manually reviewed the keywords (generated by the c-TF-IDF method) that represented the topics and examined a sample of tweets under each topic. Two authors conducted a discussion for each clustered topic and assigned it to one of the nine predetermined categories. Topics that fell under the "(M) Miscellaneous" category, which did not pertain to education settings, were excluded from subsequent analysis.

**Figure 4(a)** shows the manual classification results of the 200 topics returned by BERTopic. Using this classification, we analyzed the distribution of the original negative-sentiment 16,011 tweets, as presented in **Figure 4(b)**. Our analysis identified five frequently discussed concerns among Twitter users, including "(A) Academic integrity," "(I) Impact on learning outcomes and skill development," "(L) Limitation of capabilities," "(P) Policy and social concerns," and "(W) Workforce challenges."

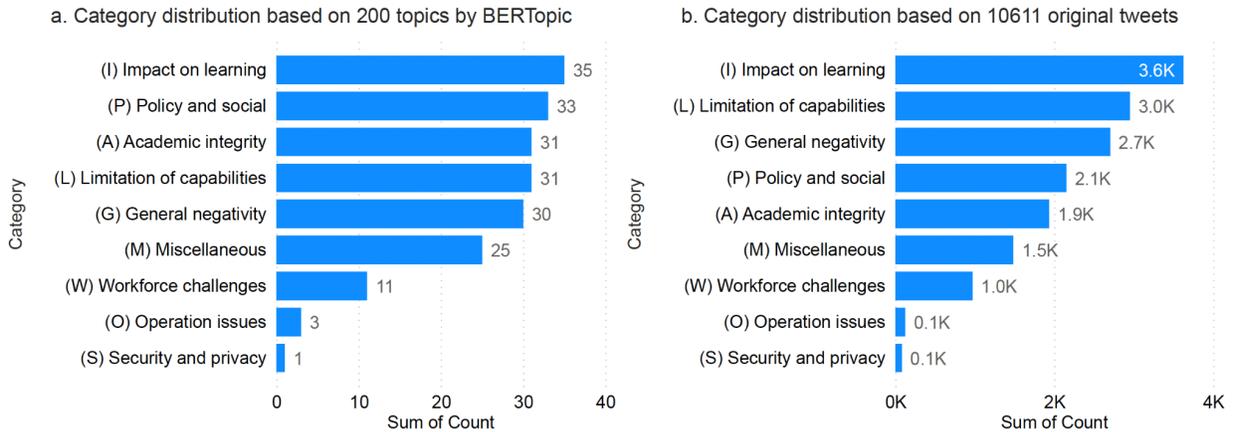

**Figure 4**. The distribution of categories based on (a) 200 topics clustered by BERTopic and (b) 10611 original negative-sentiment tweets.

We sought to examine the temporal patterns of concerns surrounding ChatGPT in education. **Figure 5** illustrates the temporal distribution of negative tweets related to these concerns. At the beginning of the study period, a surge of concerns arose regarding the potential disruption that ChatGPT could bring to education. There were also concerns about its impact on learning and skill development. In early January, when the use of ChatGPT was blocked in all New York City schools (**Figure 3**), a discussion about policy and social concerns related to ChatGPT in education emerged (i.e., could ChatGPT be used in classrooms or exams?). When ChatGPT passed MBA and law exams (**Figure 3**), Twitter users expressed concerns about its impact on the workforce, which could potentially diminish the value of education. Later in April, when tech leaders called for a pause on AI, discussions shifted towards the potential limitations of generative AI's capabilities. Our findings suggest that concerns surrounding ChatGPT in education could shift over time, with different issues taking precedence at different times depending on policies enacted and capabilities identified.

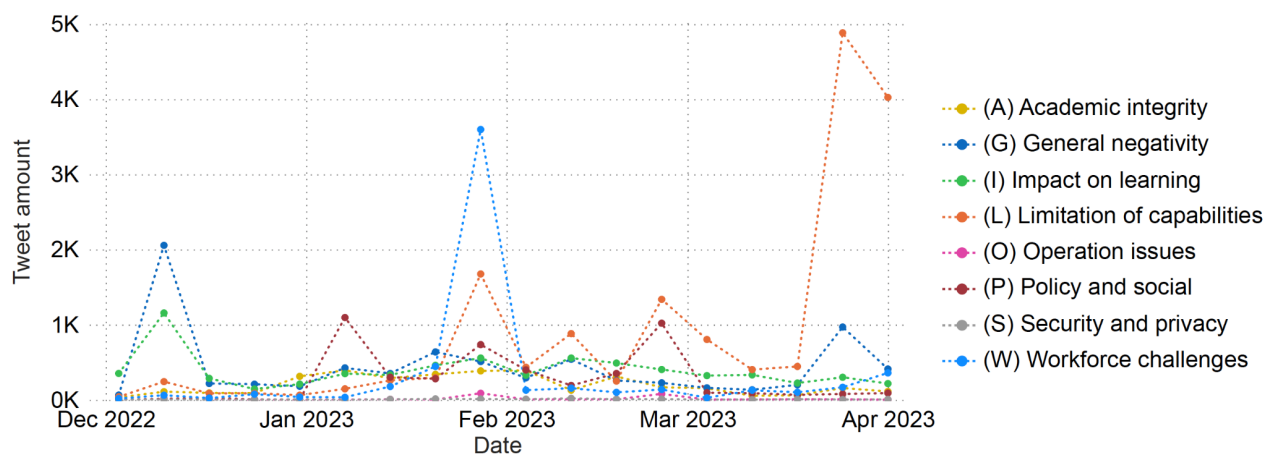

**Figure 5**. The trends of categories based on all the 49528 negative tweets.

**Table 4** provides details about each category, including its keywords and representative tweets. This data provides detail about the concerns and worries Twitter users expressed under each general category of concern.

Table 4. Representative topics and tweet examples within each category.

| Category | Topic No. | Keywords | Representative Tweet |
|---|---|---|---|
| (A) Academic integrity | 26 | ai - cheat - cheating - students - concerns - using - increase - passes - head on - over | Concerns over students using AI to cheat increase after ChatGPT passes Wharton MBA exam. |
| | 48 | plagiarism - tool - decide - whether - engine - ai - rethink - chatbot - professors - princeton | ChatGPT Is Making Universities Rethink Plagiarism. Students and professors can't decide whether the AI chatbot is a research tool, or a cheating engine. #edtech #ILoveEdTech #ImFutureReady #elearning #AI |
| | 137 | florida - scandal - erupts - elite - program - high - accuses - inside - cheating - school | ChatGPT cheating scandal erupts inside elite program at Florida high school. |
| | 152 | alert - inevitable - release - after - teachers - cheating - are - on - for - of | Teachers are on alert for inevitable cheating after release of ChatGPT. |
| (G) General negativity | 8 | education - system - we - our - world - teachers - are - will - already - change | ChatGPT is the nail in the coffin for the current Education System. RIP. #AI #Education |
| | 20 | education - system - will - pandemic - disrupt - be - we - the - already - going | ChatGPT will be truly disruptive for education. |
| | 108 | assessment - assessments - compliance - measure - deny - lazy - rethink - daughter - threat - time | Time to rethink higher education assessments. ChatGPT is disrupting everything including assessments. |

| Category | ID | Keywords | Example |
|---|---|---|---|
| (I) Impact on learning outcomes and skill development | 2 | code - coding - python - programming - learning - learn - it - me - class - ask | @scrumtuous most administrators are too worried about the ChatGPT threat to consider teaching real skills like coding. |
| | 51 | essay - essays - write - tradition - center - humanistic - writing - undergraduate - ground - student | The essay, in particular the undergraduate essay, has been the center of humanistic pedagogy for generations. It is the way we teach children how to research, think, and write. That entire tradition is about to be disrupted from the ground up. |
| | 92 | robs - motivation - write - writing - themselves - papers - term - prowess - technical - short | How #ChatGPT robs students of motivation to write and think for themselves. |
| | 128 | critical - thinking - artists - threat - panic - provider - eit - education - beans - backgrounds | If the fucking curriculum you teach can be solved by ChatGPT then you're not teaching critical thinking anyway! |
| (L) Limitation of capabilities | 33 | sources - academic - papers - source - articles - review - cite - research - references - up | @jtLOL Based off my experiments with ChatGPT, it's A LOT like journalists. It will literally make shit up. I asked questions for which I already knew the answers and it was not only wrong, but when I asked for sources, it threw back books; academic papers that did not exist. |
| | 40 | machine - learning - months - model - tuned - inept - dimwit - march - clown - limitations | In case you were wondering, ChatGPT is not a very good machine learning engineer. |
| | 70 | answers - answer - wrong - correct - mistakes - question - questions - immediately - but - errors | ChatGPT is like a lazy student who has wrong answers but always has the answers. |
| | 99 | fails - civil - exam - failed - india - competitive - service - biggest - exceeds - ka | ChatGPT fails India civil service exam. |
| (O) Operation and management issues | 109 | money - month - free - tutors - plus - evil - you - cost - your - not | ChatGPT being $42/month is bad for all. Maybe not so for westerners but for 3rd world countries, IT startups and especially for students learning new skills is just way too high!! #ChatGTP @OpenAI @sama |
| | 129 | fudan - platform - crashes - team - chatgpt style - launch - hours - apologises - after - china | China Fudan University team apologises after ChatGPT-style platform crashes hours after launch. #ChatGPT #China |
| | 164 | homework - ate - do - my - need - godsend - fun fact - hogging - dbq - tempted | Why is ChatGPT at capacity the moment that I legitimately need help with my homework |
| (P) Policy and social concerns | 58 | chatbot - inaccessible - impacts - devices - networks - negative - spokesperson - york - concerns - banned | NYC Bans Students and Teachers from Using ChatGPT. |

|  | 65 | vanderbilt - email - shooting - university - michigan - apologizes - msu - state - mass - sick | Vanderbilt University apologizes for using ChatGPT to write mass-shooting email. |
|  | 111 | conference - tools - papers - language - verge - top - bans - banned - ai - conferences | Top AI conference bans use of ChatGPT and AI language tools to write academic papers. |
|  | 158 | hong - kong - coursework - temporarily - bans - ai based - temporary - metropolitan - ban - issued | University of Hong Kong temporarily bans students from using ChatGPT, other AI-based tools for coursework. |
| (S) Security and privacy | 84 | crypto - scam - people - secrets - laid off - browses - scammers - money - engineer - scams | Will ChatGPT help hackers or defenders more?<br>Security researcher Juan Andres Guerrero-Saade has already thrown out his teaching syllabus on nation-state hackers, saying the chatbot revolutionizes painstaking efforts to uncover secrets and sorcerers behind malicious source code. |
| (W) Workforce challenges | 35 | jobs - replace - collar - class - unemployment - ke - will - useless - workers - disruptive | @MarioNawfal ChatGPT is going to kill a lot of middle class jobs. |
|  | 39 | ai - jobs - robots - education - will - spinning - workers - collar - replaced - be | This will be the first generation to ask their teacher "When are we ever going to use this?" and their teacher really not have a legitimate response.<br><br>How must education need to evolve when more than 50% of the jobs are replaced by AI in under a decade? #chatgpt #openai #society |
|  | 41 | mba - wharton - exam - pass - passes - passed - business - mbas - professor - passing | ChatGPT can already pass a Wharton business exam in case you wasted years getting a degree in a society where you need to earn money to live. |
|  | 177 | happen - on the job - white collar - discussed - react - market - labor - workers - massive - training | How will the labor market react? What will happen to on-the-job training for white-collar workers? What will happen to the massive education system as a whole? Is any of this being discussed at all? |

Under the "(G) General negativity" category, Twitter users expressed general concerns about the potential risks of ChatGPT in education. For example, some users (e.g., Topic 20 in **Table 4**) highlighted how this technology could disrupt the education system. Others worried that AI could ultimately replace the education system (e.g., Topics 8 in **Table 4**), prompting educators to reconsider assessment (e.g., Topic 108 in **Table 4**).

One specific concern addressed by Twitter communities is ChatGPT's impact on "(A) Academic integrity" in education. Users reported multiple instances of cheating with ChatGPT in academics (e.g.,

Topics 26 in **Table 4**). Concerns also existed about how ChatGPT and related AI tools could undermine the credibility of online exams and assignments, as the risk of cheating increased with the proliferation of these technologies (e.g., Topics 26 and 152 in **Table 4**). Moreover, the lack of understanding about the acceptable and unethical use of ChatGPT in an educational context, may further exacerbate these concerns (e.g., Topic 48 in **Table 4**).

In addition, we observed that the category of "(I) Impact on learning outcomes and skill development" was widely discussed within Twitter communities (**Figure 4**). Some users cautioned against using it to plagiarize or complete assignments, such as writing essays (e.g., Topics 51 in **Table 4**) or coding assignments (e.g., Topic 2 in **Table 4**). Using generative AI instead of writing prose and code one's self could undermine important abilities, such as critical thinking, communication, and problem-solving (e.g., Topics 92 and 128 in **Table 4**), raising concerns about the potential impact of ChatGPT on skill development.

Another widely discussed concern is the limitation of ChatGPT's generated information or understanding of human interactions, as illustrated in the category of "(L) Limitation of capabilities." Many users reported that ChatGPT generated nonsensical or false information (e.g., Topics 33 and 70 in **Table 4**). In particular, ChatGPT could generate resources that did not exist (Topics 33 in **Table 4**). Others doubted about ChatGPT's effectiveness as a resource for learning certain skills, such as machine learning (e.g., Topic 40 in **Table 4**) or civil service (e.g., Topic 99 in **Table 4**).

Our analysis further revealed two significant concerns expressed by Twitter users. The first one falls under the category of "(P) Policy and Social concerns." Many academic institutions have implemented policies to regulate the use of ChatGPT due to the fear of cheating or impact on learning. For instance, New York City announced the prohibition of ChatGPT in schools (e.g., Topic 58 in **Table 4**). Similarly, Hong Kong universities imposed bans on the use of ChatGPT (Topic 158 in **Table 4**). Within this category, our analysis highlights that Twitter communities engaged in extensive discussions on the ethical and moral implications of using AI-generated content in sensitive situations. For instance, one discussion revolved around whether it is appropriate to use ChatGPT to write a consolation letter to victims of mass shootings (e.g., Topic 65 in **Table 4**).

The second one falls under the category of "(W) Workforce challenges." ChatGPT's potential to offer virtual intelligent tutoring services and provide students with feedback has led to debates among educators about the possibility of being supplanted by AI chatbots (e.g., Topic 39 in **Table 4**). This concern was not limited to the education sector, as users in other fields also worried that ChatGPT could devalue their job training, such as MBA training (e.g., Topic 41 in **Table 4**). Others were concerned about ChatGPT's potential to replace middle-class jobs or white-collar jobs (e.g., Topics 35 and 177 in **Table 4**).

Last, our topic analysis reveals two additional concerns that received less attention but are still significant. One concern is related to "(O) Operation and management issues," while the other concern is "(S) Security and privacy." Concerns surrounding the operation and management mainly include the cost of premium service, stability, and capacity limitations (e.g., Topics 109, 129, and 164 in **Table 4**). These concerns reflect the challenges of maintaining a high-quality AI service that meets the needs of users. In terms of security and privacy concerns, we identified only one topic. Some users expressed concerns about the potential inadvertent disclosure of personal information and chat histories. However, we also observed a few tweets that fall into other categories but imply security and privacy issues, which is discussed in the **Limitations** section.

## 4.3. Social network analysis

As discussed in **Section 3.4**, we examined two types of relationships on Twitter: mentions and retweets. The network of mentions helps determine who was frequently implicated in the conversation and therefore potentially responsible for informing policies around ChatGPT. The network of retweets helps identify who led or disseminated the concerns. These insights help clarify the sources and information dissemination on Twitter. We utilized NodeXL to generate the networks of mentions and retweets, as presented in **Figures 6** and **7**, respectively. **Table 5** summarizes the statistics for the social networks of mentions and retweets. Explanations for each network metric are listed below.

- Vertices – Twitter users included in the network.
- Edges – Connections between two Twitter users, such as retweet and mention.
- Duplicated edges – One user mentions or retweets another user multiple times.
- Self-loops – Users mention or retweet their own tweets that form self-loops.
- Connected components – A set of users in the network that are linked to each other by edges, forming clusters within the social network.
- Geodesic distance – The shortest path between two Twitter users, measured by the least number of edges connecting them in the network.

**Table 5**. Social network statistics for the networks of mentions and retweets.

| Network metric | Network of mentions | Network of retweets |
|---|---|---|
| Network type | Directed | Directed |
| Vertices (i.e., Twitter users) | 11,478 | 32,806 |
| Total edges | 12,264 | 33,517 |
| Duplicated edges | 980 | 727 |
| Unique edges | 11,284 | 32,790 |
| Connected components | 2349 | 1499 |
| Self-loops | 155 | 199 |
| Max. geodesic distance | 26 | 21 |
| Avg. geodesic distance | 8.267 | 4.925 |

The network of mentions (**Figure 6**) comprises 11,478 vertices (i.e., Twitter users who either mentioned or were mentioned by other users), 12,264 edges, and 2,349 connected components, which are clusters within the network. The average distance between any two vertices, also known as the geodesic distance, is 8.27. In contrast, the network of retweets (**Figure 7**) comprises 32,806 vertices (i.e., Twitter users who either retweeted or were retweeted by other users), 33,517 edges, and 1,499 connected components, with an average geodesic distance of 4.93. Overall, the network of mentions has fewer users, edges, and connected components but a greater geodesic distance than the network of retweets. It is also less dense than the network of retweets (smaller geodesic distance) and has fewer clusters with a single user dominating the cluster. This suggests that there are fewer interactions between users in the network of mentions, potentially because those who were mentioned did not respond to those who mentioned them.

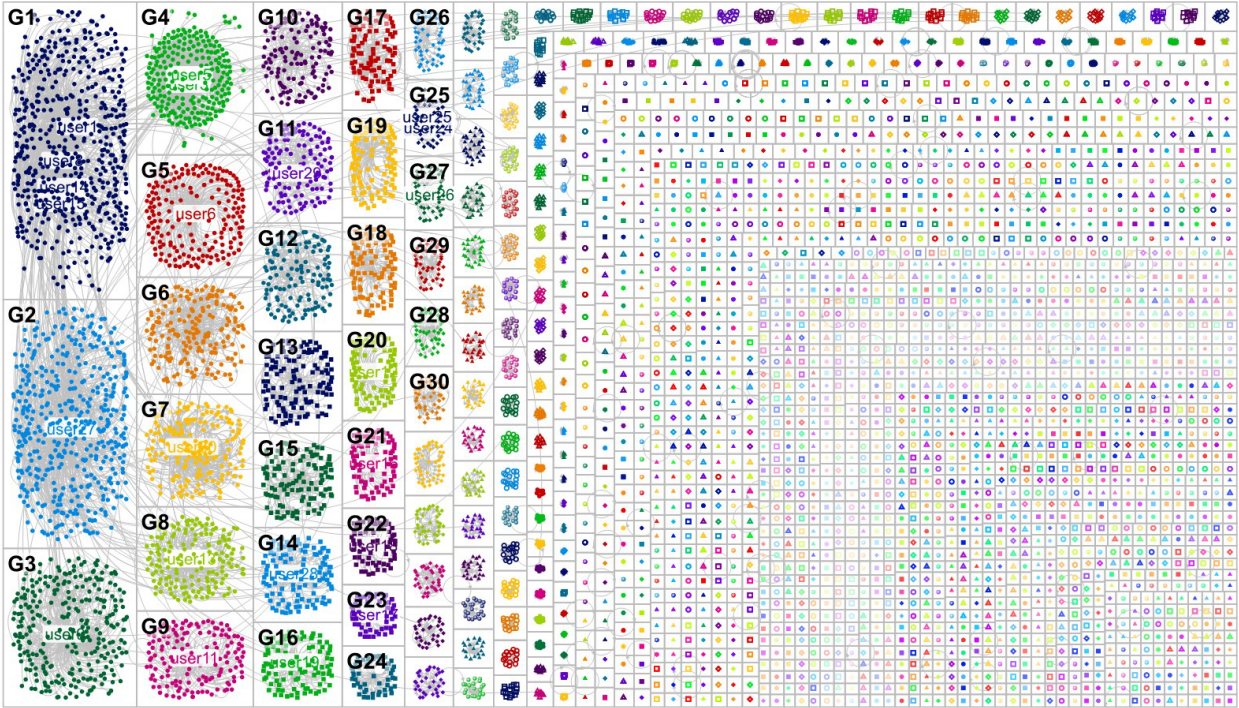

**Figure 6**. The social network of mentions.

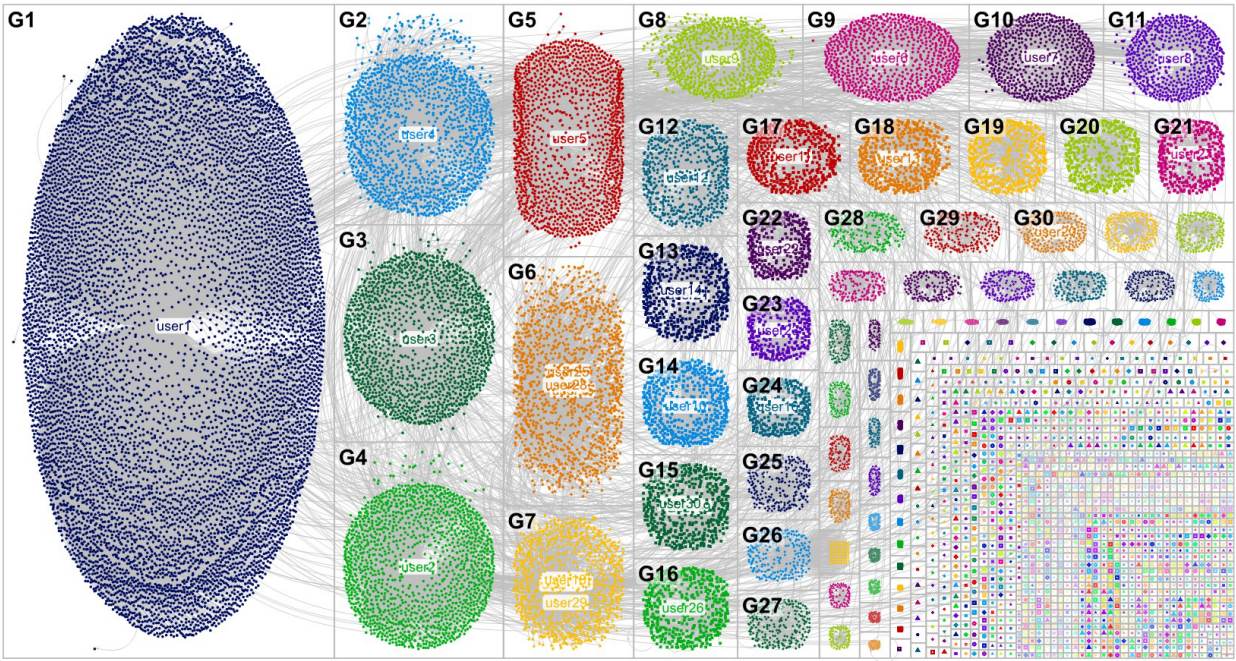

**Figure 7**. The social network of retweets.

There are several noteworthy observations regarding the users. First, a few users appear to be "centered" within the cluster, surrounded by a large number of users. These "centered" users possess the highest in-degree centrality, as they were often retweeted or mentioned by other participants in the conversation.

In some clusters, such as user 1, user 4, user 3, and user 2 in G1, G2, G3, and G4, respectively, a single user may hold complete control over the entire cluster (See **Figure 7**). In other clusters, several users co-locate in the cluster, as illustrated by user 1, user 8, user 14, and user 15 in G1 within **Figure 6**, suggesting that their tweets share common outreach and responses. We also observed that users within the network of retweets are more concentrated in comparison to those within the network of mentions, as denoted by a higher ratio of connected components divided by vertices and a greater geodesic distance. This implies that a few users' tweets were repeatedly retweeted within the community, whereas users' mentions were more diffuse. Nevertheless, many users were only mentioned or retweeted once within both networks.

**Tables 6** and **7** listed the top 30 users who have the highest in-degree centrality within the network. To provide more insights into these users, we examined two attributes: verification status and account type. Verification status indicates whether Twitter has verified an account as belonging to a public figure or organization of public interest, such as government agencies, politicians, journalists, media personalities, and other influential individuals. The second attribute is account type, which was manually determined based on our judgment of users' account description. We classified users into either organizational or individual accounts with their specialized areas, as listed below. Due to privacy considerations, we replaced usernames with "user#" in the tables.

- Organizational accounts:
    - org_gov – government agencies
    - org_tech – tech companies
    - org_media – news media
    - org_edu – education institutions
    - org_other – other types of organizations
- Individual accounts:
    - ind_politic – politicians
    - ind_tech – tech engineers and businessmen
    - ind_media – media workers
    - ind_edu – educators
    - ind_other – other types of users

Regarding the users, we made several notes from results presented in **Tables 6** and **7**. In the network of mentions, 43% (13 out of 30) of the top-mentioned users have verified accounts, primarily from the tech industry, followed by education and media accounts. In contrast, for the network of retweets, most top users are independent influencers with non-verified accounts (25 out of 30). We also noticed a clear difference in the presence of organizational accounts between the two networks. More than half of top mentioned accounts are from organizations, whereas only one organization made it to the top users in the network of retweets. Furthermore, these top retweeted accounts are mostly from individuals engaged in education or tech. However, accounts associated with politicians and government portals rarely appear among the top users.

Regarding the categories, we found that these top users were frequently mentioned in tweets related to (L) Limitations of capabilities (users 2 to 5 in **Table 6**), (I) Impact on learning outcomes and skill development (users 6, 9, 11, and 20 to 25 in **Table 6**), (W) Workforce challenges (users 1, 10, 13, and 16 in **Table 6**), (G) General negativity (users 1, 8, 14, and 15 in **Table 6**), and (A) Academic integrity (users 17, 19, 26, and 29 in **Table 6**). However, they were comparatively less-often mentioned in other categories.

Similarly, based on categories of retweets (see **Table 7**), tweets within these five identified categories were more likely to receive attention from the community, while the other categories were not as prominent.

**Table 6**. Top 30 users based on in-degree centrality in the network of mentions.

| User | In-degree | Group | Verified | Account | Categories |
|---|---|---|---|---|---|
| User 1 | 386 | G1 | True | org_tec | W(124), G(109), I(53), L(50), A(19), P(13), O(4) |
| User 2 | 213 | G4 | False | ind_tech | L(219), I(5) |
| User 3 | 210 | G4 | True | org_tech | L(205), A(2), I(2), W(1), P(1) |
| User 4 | 207 | G4 | False | org_tech | L(204) |
| User 5 | 202 | G4 | False | org_tech | L(204) |
| User 6 | 174 | G5 | False | org_tech | I(173) |
| User 7 | 154 | G2 | True | ind_tech | L(53), G(31), I(13), A(12), W(11), P(11), O(3) |
| User 8 | 125 | G1 | True | ind_tech | G(93), L(15), P(5), I(4), W(3), O(1) |
| User 9 | 124 | G2 | True | org_media | I(114), G(2), L(1) |
| User 10 | 113 | G8 | True | ind_tech | W(65), I(16), G(10), A(9), L(6), P(6), O(1) |
| User 11 | 88 | G9 | False | org_media | I(83), G(1), L(1) |
| User 12 | 80 | G20 | True | ind_edu | L(80) |
| User 13 | 71 | G8 | False | ind_edu | W(64), I(4), L(2), A(1) |
| User 14 | 71 | G1 | True | ind_tech | G(70) |
| User 15 | 70 | G1 | False | ind_tech | G(70) |
| User 16 | 67 | G21 | False | ind_edu | W(68) |
| User 17 | 51 | G23 | True | org_media | A(50) |
| User 18 | 49 | G22 | False | ind_other | L(16), G(10), W(10), I(5), A(3), P(2) |
| User 19 | 47 | G16 | True | org_media | A(40), I(2), L(20), W(2), P(2), G(1) |
| User 20 | 42 | G25 | True | ind_edu | I(46) |
| User 21 | 42 | G25 | False | ind_edu | I(46) |
| User 22 | 42 | G25 | False | ind_edu | I(46) |
| User 23 | 42 | G25 | False | ind_edu | I(46) |
| User 24 | 42 | G25 | False | ind_edu | I(46) |
| User 25 | 42 | G25 | False | ind_edu | I(43) |
| User 26 | 41 | G27 | True | org_media | A(41) |
| User 27 | 40 | G2 | True | org_other | G(14), W(9), L(7), I(5), A(2), P(1) |
| User 28 | 36 | G14 | False | org_tech | L(20), G(19), I(8), A(6), W(6), P(4) |
| User 29 | 33 | G11 | False | ind_tech | A(31), G(1), I(1) |
| User 30 | 29 | G7 | False | ind_tech | P(22), I(12), A(4), G(1) |

(Notes: (1) The "Category" column in **Table 6** implies the number of mentions of a user. (2) It should be noted that the sum of mentions is not equal to the in-degree centrality in **Table 6**. This is because in-degree centrality only counts unique edges between two users (unique edges are mentions by one other account; even if account A mentions account B twice, the in-degree for account B is 1), and includes tweets labeled as "(M) Miscellaneous.")

Table 7. Top 30 users based on in-degree centrality in the network of retweets.

| User | In-degree | Group | Verified | Account | Categories |
| --- | --- | --- | --- | --- | --- |
| User 1 | 9042 | G1 | False | ind_edu | I(4), L(4), G(1), W(1), A(1) |
| User 2 | 1562 | G4 | False | ind_tech | W(1) |
| User 3 | 1489 | G3 | False | ind_tech | G(1) |
| User 4 | 1295 | G2 | False | ind_edu | W(1), I(1) |
| User 5 | 1287 | G5 | False | ind_edu | L(1) |
| User 6 | 725 | G9 | False | ind_other | G(1) |
| User 7 | 575 | G10 | False | ind_other | G(2), W(1) |
| User 8 | 520 | G11 | False | ind_tech | W(2) |
| User 9 | 460 | G8 | True | ind_politic | P(1) |
| User 10 | 354 | G14 | False | ind_other | L(1), I(1) |
| User 11 | 343 | G17 | False | ind_edu | L(1) |
| User 12 | 285 | G12 | True | ind_edu | I(1), G(1), A(1) |
| User 13 | 281 | G18 | True | ind_media | I(1) |
| User 14 | 224 | G13 | False | org_media | L(1) |
| User 15 | 211 | G6 | False | ind_edu | A(1) |
| User 16 | 206 | G24 | False | ind_edu | L(1) |
| User 17 | 198 | G7 | False | ind_edu | I(2) |
| User 18 | 189 | G7 | False | ind_edu | G(1) |
| User 19 | 186 | G15 | True | ind_edu | G(3), I(2), A(1), L(1), P(1) |
| User 20 | 180 | G30 | False | ind_tech | I(1), P(1) |
| User 21 | 179 | G13 | False | ind_tech | P(1) |
| User 22 | 156 | G22 | False | ind_tech | P(1) |
| User 23 | 154 | G23 | False | ind_tech | M(1) |
| User 24 | 129 | G7 | False | ind_edu | G(1) |
| User 25 | 112 | G6 | False | ind_tech | G(1) |
| User 26 | 104 | G16 | False | ind_edu | L(1) |
| User 27 | 103 | G21 | False | ind_tech | W(1), P(1) |
| User 28 | 97 | G6 | False | ind_edu | W(1) |
| User 29 | 96 | G7 | False | ind_edu | G(1) |
| User 30 | 93 | G15 | True | ind_edu | I(1) |

(Note: the "Category" column in **Table 7** implies the number of tweets that the user posted. User 23 only has one tweet fallen into the "(M) Miscellaneous" category.)

In addition to analyzing the network of the top 30 mentioned users, as shown in **Table 6**, we sought to examine user patterns on a broader scale. To achieve this, we conducted further investigations on accounts with an in-degree larger than 12, resulting in a total of 105 users. Upon manually reviewing their profiles, there were four users who had either deactivated their accounts or had posted zero tweets. As a result, we included 101 users to explore the patterns from mentioned users, as presented in **Figure 8**.

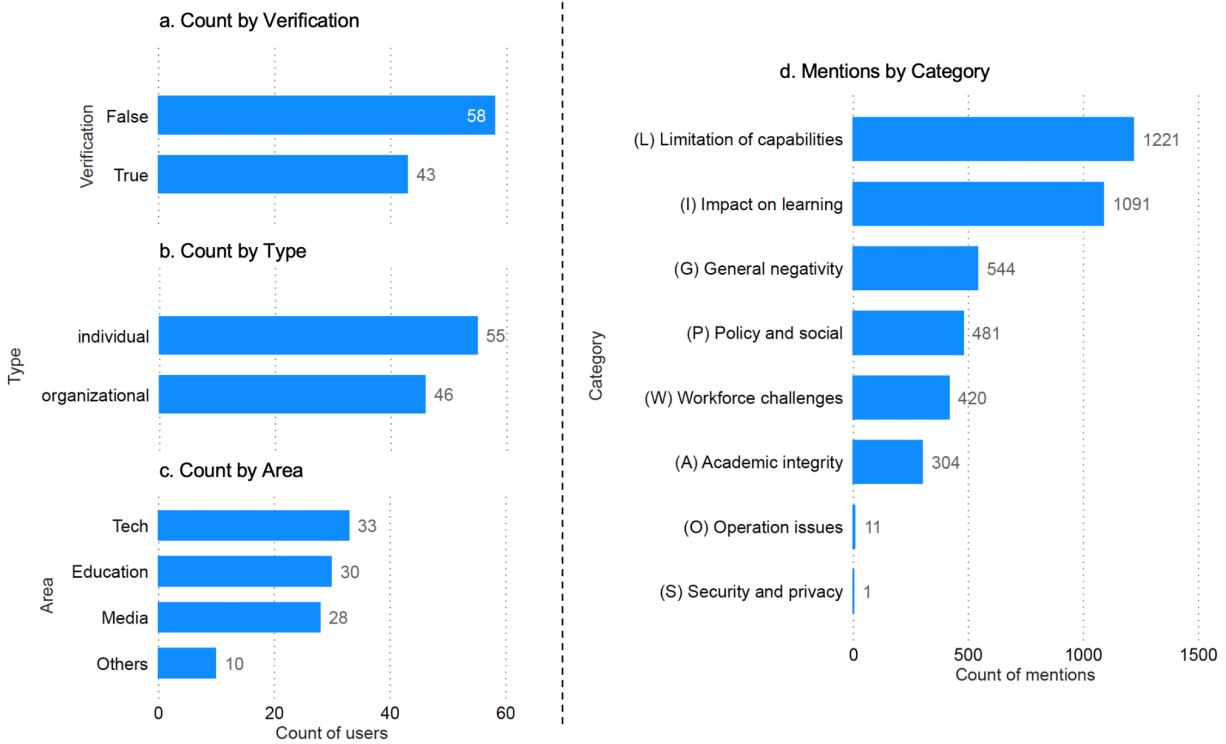

**Figure 8.** Distribution of top 101 mentioned users in the network. (a) Count by verification. (b) Count by type. (c) Count by area. (d) Mentions by category.

Upon analyzing **Figure 8(a)-(c)**, we discovered that more than half of the top mentioned users (58 out of 101) were non-verified and individual accounts (55 out of 101). In terms of account area, there was an almost equal distribution of top mentioned users from the tech, education, and media sectors. However, government portals and politicians, who we expected to be implicated in the discussion, were rarely mentioned in this list. Further manual examination of these accounts revealed that many of these tech accounts are from big tech companies such as OpenAI, Google, and Microsoft or influential figures like Elon Musk and Sam Altman. The education accounts are mainly university professors, and the media accounts are from news agencies, such as the *New York Times* and the *Washington Post*.

Upon examination of **Figure 8(d)**, we identified that the top users were predominantly mentioned in two categories: (L) Limitation of capabilities (mentioned 1221 times) and (I) Impact on learning outcomes and skill development (mentioned 1091 times). After manual examination, we found that the most frequently mentioned three accounts for both categories are from the tech area. This pattern persists across other categories, with the exception of (A) academic integrity, where media accounts take center stage.

# 5. Discussion

As an emerging technology, ChatGPT has the potential to bring significant opportunities to the education sector, but it also poses certain risks. To better understand how the public perceives these risks and concerns, we conducted a study to identify concerns about the ChatGPT application in education as expressed in social media. Specifically, we chose a crowdsourcing approach with Twitter data, aiming to (1) tap into a diverse range of perspectives from the public and (2) identify users who were frequently implicated and who communicated the concerns in the conversation. Our study sheds light on the public's perceptions of key issues and considerations that need to be addressed for the responsible deployment of AI applications in education. Below are the key findings.

**1. Twitter users expressed generally positive attitudes towards using ChatGPT in education.** The sentiment analysis indicates that a significant majority of users hold a positive view of ChatGPT's potential applications in education. Despite the relatively recent availability of this technology, many Twitter users appear to be enthusiastic about the ways in which it could improve the learning experience for students and teachers alike. This observation aligns with Tlili et al. (2023), showing that Twitter users showed a generally positive sentiment in the month following the release of ChatGPT. This consistency suggests that Twitter users continued to view the application of ChatGPT in education positively.

In addition, our findings suggest that major events can influence public perception of ChatGPT's potential benefits and risks. For instance, when ChatGPT passed exams typically taken by human professionals, such as MBA and law exams (Kelly, 2023; Rosenblatt, 2023), some users voiced concerns about its impact on the job market and the value of educational training. We also observed variability in sentiment after updates from the model creators and when tech industry professionals made public comments about generative AI.

**2. Twitter users mainly focused on five specific concerns.** In response to the first research question, our findings reveal that Twitter users mainly discussed five specific concerns: "academic integrity," "impact on learning outcomes and skill development," "limitation of capabilities," "policy and social concerns," and "workforce challenges." These concerns echo prior work, which explains expert opinions (AlAfnan et al., 2023; Mhlanga, 2023; Mintz, 2023; Qadir, 2022; Sok & Heng, 2023). Twitter users identified ChatGPT's potential risks to academic integrity, writing and coding skills, limited capabilities, and certain job markets.

Twitter users also expressed their unique perspectives based on their experiences with the technology. For instance, many users reported instances of ChatGPT generating misinformation, false references, and incorrect answers, which is also consistent with literature we reviewed (Baidoo-Anu & Owusu Ansah, 2023; Qadir, 2022). Other users noted that ChatGPT can be at capacity or break down, leading to frustration and limitations in usage. Users also expressed concerns about the limitations of ChatGPT in specialized areas, such as machine learning or civil service exams, where it may provide limited or insufficient information.

We observed discussion on Twitter regarding the banning of ChatGPT in schools and universities due to concerns about its impact on academic integrity and learning outcomes. Last, many users expressed concerns about negative social consequences that could arise from using ChatGPT, such as using it to draft insincere or impersonal letters in sensitive situations (Korn, 2023).

**3. Tech, education, and media users were highly implicated in the conversation**. In response to the second research question, our analysis of social networks first revealed that Twitter users mentioned others from the tech, education, and media sectors. Specifically, we observed that accounts from big tech companies or influential individuals in the tech sector were often mentioned. Professors were the key players in the education sector, and popular news agencies led the media sector. However, we found no significant difference in the level of influence between verified or unverified accounts, nor did account type (i.e., individual or organizational) make a significant impact on the conversation's direction. Politicians and governmental agents were barely mentioned in the conversation. Together these findings indicate that Twitter is not actively discussing regulation or government oversight of generative AI. Instead, Twitter is trying to understand the technology (i.e., by engaging academic experts), to follow related news, and to keep up with tech industry conversations.

**4. Education and tech individual users drove the distribution of concerns.** Our analysis of social networks also revealed that individual users from education and tech sectors played a vital role in driving the discussion about concerns. Specifically, we identified 30 users whose tweets about concerns were retweeted most often. Among them, a substantial proportion of the top users who drove the conversation on ChatGPT's use in education belonged to the education sector, specifically professors. This observation indicates that professors hold significant influence in shaping the discourse around education, and their concerns on ChatGPT's application in education were well-received by the public. We recognize that a fundamental role of academia is to question and critique technological advances, and professors may be over-represented in the population of users who express concerns about ChatGPT.

## 5.1. Practical implications

Our study suggests several practical implications for various stakeholders, including policymakers, tech companies and individuals, educators, and media agents. In particular, these practical implications are from the general public's perspective on social media. These implications center on the need for a broader discussion on the integration of AI with education, with a focus on preparing individuals to work alongside these technologies effectively.

**1. Policymakers should take a more proactive role in the conversation about AI in education**. Our findings reveal that government officials and politicians rarely participated in the discussion, as evidenced by the low number of accounts implicated (the network of mentions) or distributing the concerns (the network of retweets). However, on April 11, 2023, the Biden administration called for public comments on potential accountability measures for AI systems (Tracy, 2023), reflecting a growing interest in AI regulation. In addition, the National Telecommunications and Information Administration agency advised the White House to take actions on telecommunications and information policy (Shepardson & Bartz, 2023). According to

Based on the guidance outlined by Carnegie Mellon University in 2020 (CMU Block Center for Technology and Society, 2020), we suggest that government agencies should take the role of policymakers in shaping AI policies. Specifically, the paper suggested that the federal government should update the regulatory guidance of existing agencies, such as Security and Exchange Commission (SEC), Consumer Financial Protection Bureau (CFPB), Food and Drug Administration (FDA) to ensure AI-assisted decision-making. Regarding the generative AI policy, a recent report indicates that the US government has taken actions in regulating generative AI primarily through the Federal Trade Commission (FTC) (Ryan-Mosley, 2023).

In the educational settings, we recommend that policymakers should take a more proactive role in the conversation of AI policy. In particular, they could address the implications of generative AI models in education and develop clear guidelines for their use. Such guidelines should also consider the potential social impacts and academic integrity concerns associated with AI-generated documents, such as "policy and social concerns" and "academic integrity," that emerged in our sample. Finally, we urge policymakers to carefully evaluate how AI models may challenge the current education system, particularly in terms of assessment methods.

**2. Tech companies should take responsibility for improving the capabilities and collaborating with policymakers and educators to determine policies.** Our analysis of the social network revealed that both organizational and individual tech accounts were frequently mentioned. These accounts include prominent tech companies such as Google, Microsoft, OpenAI, and influential individuals in the tech industry such as Elon Musk and Sam Altman. We observed that these tech accounts were mostly referenced in the "limitation of capabilities" category. As a result, we recommend that these tech companies focus on improving their AI models' performance and developing responsible and effective educational tools that incorporate AI.

In addition, given that these tech accounts were frequently mentioned across all categories, we suggest that they collaborate with educators and policymakers to ensure the safe and ethical use of AI in education. We also noticed that some users have encountered issues when using ChatGPT, such as breakdowns or limitations in capabilities, and that some users cannot afford to use the premium version. Tech companies may also need to address operational and management issues to provide users with a more seamless experience when using AI products.

Their prominence in the conversation about concerns suggests that Twitter users see tech companies and leaders as stakeholders. Twitter users have recommendations for improvement and valid concerns that tech companies can leverage to improve AI and its utility. They also have a responsibility to work with policymakers, within government and civil society, to communicate limitations and appropriate use of generative AI.

**3. Educators should voice their concerns and leverage AI as a learning tool in the classroom.** As the main users of ChatGPT in education, we recommend that educators share their concerns about AI in education on social media. Our analysis of the retweet network indicates that professors' views were widely retweeted by Twitter users, implying that the public trusted, or at least considered, their opinions on the topic. We also observed that discussions around academic integrity and AI policy were prevalent in Twitter communities.

Rather than prohibiting the use of AI tools in education, it might be more beneficial to incorporate them into conventional teaching methodologies and revise guidelines regarding academic misconduct. Our suggestion aligns with an opinion article published in the *New York Times* (Roose, 2023). Our analysis reveals that many individuals perceive ChatGPT as a tool that has the potential to augment their learning experiences, similar to calculators or other educational aids. In addition, we suggest that educators should focus on guiding students in the appropriate use of the AI tools rather than trying to ban them outright. Specifically, students should be taught how to navigate the technology and apply critical thinking and problem-solving skills to their use (Sun & Hoelscher, 2023).

**4. Media agencies should play an important role in supervision.** Our findings show that media accounts were frequently mentioned in the conversation, and sentiment changes were associated with reported events. We recommend that media agencies provide accurate and impartial coverage of ChatGPT, including its capabilities, limitations, and potential applications in education. Topics such as whether AI could be used to create documents in sensitive situations or if it has a place in the classroom should be carefully examined and reported on by media agents. Doing so can not only help to foster informed public discussion but also supervise responsible and ethical development of AI in education.

## 5.2. Limitations and future work

While our data-driven crowdsourcing approach to characterizing public response to generative AI in education has provided valuable insights, it's important to note several limitations. First, our data collection spans only four months after the release of the ChatGPT application. It's possible that some people's attitudes towards ChatGPT may change with a longer period of usage, particularly as new GPT4 models and policies for integrating AI models in education emerge. Moreover, there could be new risks associated with using Generative AI models in education that arise over time. In response to this limitation, an area of ongoing and future work could involve the continued collection of social media data to monitor emerging sentiment and concerns among the public.

Another limitation comes from sentiment analysis. First, the accuracy of the sentiment analysis is contingent on the capability of the RoBERTa-base model. Given that we did not train the model specifically for sentiment analysis in this context, it is possible that the model may incorrectly detect sentiment in some cases. Last, the sentiment score was calculated based on tweet units, but it's worth noting that a tweet may include conflicting sentiments in different parts. To address this limitation, one future study could investigate other sentiment detection models, such as BERTweet (Nguyen et al., 2020), and compare their performance.

There are two limitations associated with the topic modeling approach employed (BERTopic). First, the assumption that each document contains only one topic does not always hold in tweets. As a result, BERTopic can return suboptimal representations of documents. Second, we observed that certain clustered topics may encompass tweets pertaining to different categories, possibly because BERTopic clusters tweets based solely on their textual similarity. In particular, those clustered topics that contain a significant number of tweets are more likely to include descriptions of multiple categories. To address this limitation, future work could consider exploring other LLMs such as integrating GPT-based models (Chen et al., 2021) with the topic modeling process.

While crowdsourcing can help mitigate biases that may arise in open-ended surveys, it's important to acknowledge that relying on public comments about the ChatGPT application in education introduces its own biases. People who write tweets are not representative of the general public. For example, research has shown that young, educated, and urbanized individuals are more likely to post comments on social media platforms due to their familiarity with social media (Barberá & Rivero, 2015; Mislove et al., 2021). Additionally, some individuals may express their opinions on other social media platforms such as Facebook or Tiktok and may use videos or images to convey their attitudes towards ChatGPT applications. These factors could potentially affect the quality of the data preparation and introduce biases into the results. A prior study shows some potential of using TikTok videos to capture people's opinions regarding the use of ChatGPT (Haensch et al., 2023). Therefore, another possible avenue for future research could include data from other social media platforms, such as Facebook or TikTok, or from more representative sources such as probability sample surveys. Each platform and method has potential trade-offs, and we acknowledge these limitations of the Twitter sample.

# 6. Conclusions

Generative AI models have the potential to revolutionize education but also present risks that must be carefully considered. The emergence of ChatGPT has sparked a significant amount of discussion on social media about its potential applications in education. To contribute to this conversation, our study uses a crowdsourcing approach to identify concerns by analyzing discourse on Twitter. Specifically, we employ BERT-based sentiment and topic modeling techniques to identify concerns related to the use of ChatGPT and use social network theory to identify key accounts frequently implicated in the discussion.

The sentiment analysis indicates that Twitter users have an overall positive attitude towards the use of ChatGPT in education. However, we note that sentiment changes are often associated with significant events that occur within the conversation. Our topic analysis highlights five key areas of concern that emerged from negative tweets: academic integrity, impact on learning outcomes and skill development, limitation of capabilities, policy and social concerns, and workforce challenges. Our social network analysis shows that users from the fields of tech, education, and media were highly implicated in the conversation, while education and tech individual users played a crucial role in leading the diffusion of concerns to broader audiences.

Taken together, our discourse analysis underscores the urgent need for collaboration among policymakers, tech companies and individuals, educators, and media agencies to establish guidelines for the use of AI in education. While generative AI models offer significant opportunities for enhancing learning, we must address the identified concerns and risks in a responsible and ethical manner. By working together, we can develop effective policies and guidelines that ensure the responsible use of AI in education for the benefit of all stakeholders.

## Author contributions

- **Lingyao Li**: Conceptualization, Methodology, Data Curation, Formal Analysis, Writing - Original Draft, Writing - Review & Editing.
- **Zihui Ma**: Methodology, Data Curation, Formal Analysis, Writing - Original Draft, Writing - Review & Editing.
- **Lizhou Fan**: Methodology, Data Curation, Writing - Original Draft, Writing - Review & Editing.
- **Sanggyu Lee**: Writing - Original Draft, Writing - Review & Editing.
- **Huizi Yu**: Writing - Original Draft, Writing - Review & Editing.
- **Libby Hemphill**: Conceptualization, Writing - Review & Editing, Project Administration, Resources.

## Declaration of competing interest

The authors declare that they have no known competing interests or personal relationships that could have appeared to influence the work reported in this paper.

## Acknowledgement

This material is based upon work supported by the National Science Foundation under grant no. 1928434.

# References

Abdelghani, R., Wang, Y.-H., Yuan, X., Wang, T., Lucas, P., Sauzéon, H., & Oudeyer, P.-Y. (2022). *GPT-3-driven pedagogical agents for training children's curious question-asking skills*. https://doi.org/10.48550/ARXIV.2211.14228

Adiguzel, T., Kaya, M. H., & Cansu, F. K. (2023). Revolutionizing education with AI: Exploring the transformative potential of ChatGPT. *Contemporary Educational Technology*, *15*(3), ep429. https://doi.org/10.30935/cedtech/13152

Ahmed, Y. A., Ahmad, M. N., Ahmad, N., & Zakaria, N. H. (2019). Social media for knowledge-sharing: A systematic literature review. *Telematics and Informatics*, *37*, 72–112. https://doi.org/10.1016/j.tele.2018.01.015

AlAfnan, M. A., Samira Dishari, Marina Jovic, & Koba Lomidze. (2023). ChatGPT as an Educational Tool: Opportunities, Challenges, and Recommendations for Communication, Business Writing, and Composition Courses. *Journal of Artificial Intelligence and Technology*. https://doi.org/10.37965/jait.2023.0184

Alser, M., & Waisberg, E. (2023). Concerns with the usage of ChatGPT in Academia and Medicine: A viewpoint. *American Journal of Medicine Open*, 100036. https://doi.org/10.1016/j.ajmo.2023.100036

Atlas, S. (2023). *ChatGPT for Higher Education and Professional Development: A Guide to Conversational AI*. College of Business Faculty Publications. https://digitalcommons.uri.edu/cba_facpubs/548

Baidoo-Anu, D., & Owusu Ansah, L. (2023). Education in the Era of Generative Artificial Intelligence (AI): Understanding the Potential Benefits of ChatGPT in Promoting Teaching and Learning. *SSRN Electronic Journal*. https://doi.org/10.2139/ssrn.4337484

Barberá, P., & Rivero, G. (2015). Understanding the Political Representativeness of Twitter Users. *Social Science Computer Review*, *33*(6), 712–729. https://doi.org/10.1177/0894439314558836

Barbieri, F., Camacho-Collados, J., Espinosa Anke, L., & Neves, L. (2020). TweetEval: Unified Benchmark and Comparative Evaluation for Tweet Classification. *Findings of the Association for Computational Linguistics: EMNLP 2020*, 1644–1650. https://doi.org/10.18653/v1/2020.findings-emnlp.148

Bernius, J. P., Krusche, S., & Bruegge, B. (2022). Machine learning based feedback on textual student answers in large courses. *Computers and Education: Artificial Intelligence*, *3*, 100081. https://doi.org/10.1016/j.caeai.2022.100081

Buitinck, L., Louppe, G., Blondel, M., Pedregosa, F., Mueller, A., Grisel, O., Niculae, V., Prettenhofer, P., Gramfort, A., Grobler, J., Layton, R., Vanderplas, J., Joly, A., Holt, B., & Varoquaux, G. (2013). *API design for machine learning software: Experiences from the scikit-learn project*. https://doi.org/10.48550/ARXIV.1309.0238

Chen, M., Tworek, J., Jun, H., Yuan, Q., Pinto, H. P. de O., Kaplan, J., Edwards, H., Burda, Y., Joseph, N., Brockman, G., Ray, A., Puri, R., Krueger, G., Petrov, M., Khlaaf, H., Sastry, G., Mishkin, P., Chan, B., Gray, S., … Zaremba, W. (2021). *Evaluating Large Language Models Trained on Code*. https://doi.org/10.48550/ARXIV.2107.03374

Chen, T.-J. (2023). ChatGPT and other artificial intelligence applications speed up scientific writing. *Journal of the Chinese Medical Association*, *86*(4), 351–353. https://doi.org/10.1097/JCMA.0000000000000900

Choi, J. H., Hickman, K. E., Monahan, A., & Schwarcz, D. B. (2023). ChatGPT Goes to Law School. *SSRN Electronic Journal*. https://doi.org/10.2139/ssrn.4335905

Clauset, A., Newman, M. E. J., & Moore, C. (2004). Finding community structure in very large networks. *Physical Review E*, *70*(6), 066111. https://doi.org/10.1103/PhysRevE.70.066111

CMU Block Center for Technology and Society. (2020). *A Policy Maker's Guide to Artificial Intelligence for State and Local Governments: Reaching Safe, Effective and Equitable Scale*. Carnegie Mellon University. https://www.cmu.edu/block-center/files/andes-whitepaper-policymakers-guide.pdf

Devlin, J., Chang, M.-W., Lee, K., & Toutanova, K. (2019). BERT: Pre-training of Deep Bidirectional Transformers for Language Understanding. *Proceedings of the 2019 Conference of the North*, 4171–4186. https://doi.org/10.18653/v1/N19-1423

Dijkstra, R., Genc, Z., Kayal, S., & Kamps, J. (2022). *Reading Comprehension Quiz Generation using Generative Pre-trained Transformers*.

Dwivedi, Y. K., Kshetri, N., Hughes, L., Slade, E. L., Jeyaraj, A., Kar, A. K., Baabdullah, A. M., Koohang, A., Raghavan, V., Ahuja, M., Albanna, H., Albashrawi, M. A., Al-Busaidi, A. S., Balakrishnan, J., Barlette, Y., Basu, S., Bose, I., Brooks, L., Buhalis, D., … Wright, R. (2023). "So what if ChatGPT wrote it?" Multidisciplinary perspectives on opportunities, challenges and implications of generative conversational AI for research, practice and policy. *International Journal of Information Management*, *71*, 102642. https://doi.org/10.1016/j.ijinfomgt.2023.102642

Fan, L., Li, L., Ma, Z., Lee, S., Yu, H., & Hemphill, L. (2023). *A Bibliometric Review of Large Language Models Research from 2017 to 2023*. https://doi.org/10.48550/ARXIV.2304.02020

Fauzi, F., Tuhuteru, L., Sampe, F., Ausat, A. M. A., & Hatta, H. R. (2023). Analysing the Role of ChatGPT in Improving Student Productivity in Higher Education. *Journal on Education*, *5*(4), 14886–14891. https://doi.org/10.31004/joe.v5i4.2563

Feng, Y., Poralla, P., Dash, S., Li, K., Desai, V., & Qiu, M. (2023). *The Impact of ChatGPT on Streaming Media: A Crowdsourced and Data-Driven Analysis using Twitter and Reddit*.

Fijačko, N., Gosak, L., Štiglic, G., Picard, C. T., & John Douma, M. (2023). Can ChatGPT pass the life support exams without entering the American heart association course? *Resuscitation*, *185*, 109732. https://doi.org/10.1016/j.resuscitation.2023.109732

Firat, M. (2023). *How Chat GPT Can Transform Autodidactic Experiences and Open Education?* [Preprint]. Open Science Framework. https://doi.org/10.31219/osf.io/9ge8m

Fruchterman, T. M. J., & Reingold, E. M. (1991). Graph drawing by force-directed placement. *Software: Practice and Experience*, *21*(11), 1129–1164. https://doi.org/10.1002/spe.4380211102

Future of Life Institute. (2023, March 22). Pause Giant AI Experiments: An Open Letter. *Future of Life Institute*. https://futureoflife.org/open-letter/pause-giant-ai-experiments/

Gabajiwala, E., Mehta, P., Singh, R., & Koshy, R. (2022). Quiz Maker: Automatic Quiz Generation from Text Using NLP. In P. K. Singh, S. T. Wierzchoń, J. K. Chhabra, & S. Tanwar (Eds.), *Futuristic Trends in Networks and Computing Technologies* (Vol. 936, pp. 523–533). Springer Nature

Singapore. https://doi.org/10.1007/978-981-19-5037-7_37

Grootendorst, M. (2022). *BERTopic: Neural topic modeling with a class-based TF-IDF procedure*. https://doi.org/10.48550/ARXIV.2203.05794

Haensch, A.-C., Ball, S., Herklotz, M., & Kreuter, F. (2023). *Seeing ChatGPT Through Students' Eyes: An Analysis of TikTok Data* (arXiv:2303.05349). arXiv. http://arxiv.org/abs/2303.05349

Halaweh, M. (2023). ChatGPT in education: Strategies for responsible implementation. *Contemporary Educational Technology*, *15*(2), ep421. https://doi.org/10.30935/cedtech/13036

Haleem, A., Javaid, M., & Singh, R. P. (2022). An era of ChatGPT as a significant futuristic support tool: A study on features, abilities, and challenges. *BenchCouncil Transactions on Benchmarks, Standards and Evaluations*, *2*(4), 100089. https://doi.org/10.1016/j.tbench.2023.100089

Haque, M. U., Dharmadasa, I., Sworna, Z. T., Rajapakse, R. N., & Ahmad, H. (2022). *"I think this is the most disruptive technology": Exploring Sentiments of ChatGPT Early Adopters using Twitter Data*. https://doi.org/10.48550/ARXIV.2212.05856

Hurst, L. (2023, March 29). *"Profound risk to humanity": Experts call for halt to AI development*. Euronews. https://www.euronews.com/next/2023/03/29/profound-risk-to-humanity-elon-musk-and-steve-wozniak-join-calls-to-halt-ai-development

Iskender, A. (2023). Holy or Unholy? Interview with Open AI's ChatGPT. *European Journal of Tourism Research*, *34*, 3414. https://doi.org/10.54055/ejtr.v34i.3169

Jia, Q., Cui, J., Xiao, Y., Liu, C., Rashid, P., & Gehringer, E. F. (2021). *ALL-IN-ONE: Multi-Task Learning BERT models for Evaluating Peer Assessments* (arXiv:2110.03895). arXiv. http://arxiv.org/abs/2110.03895

Johnson, A. (2023, January 23). *ChatGPT In Schools: Here's Where It's Banned—And How It Could Potentially Help Students*. Forbes. https://www.forbes.com/sites/ariannajohnson/2023/01/18/chatgpt-in-schools-heres-where-its-banned-and-how-it-could-potentially-help-students/?sh=443643686e2c

Kasneci, E., Sessler, K., Küchemann, S., Bannert, M., Dementieva, D., Fischer, F., Gasser, U., Groh, G., Günnemann, S., Hüllermeier, E., Krusche, S., Kutyniok, G., Michaeli, T., Nerdel, C., Pfeffer, J., Poquet, O., Sailer, M., Schmidt, A., Seidel, T., … Kasneci, G. (2023). ChatGPT for good? On opportunities and challenges of large language models for education. *Learning and Individual Differences*, *103*, 102274. https://doi.org/10.1016/j.lindif.2023.102274

Kelly, S. M. (2022, December 5). *This AI chatbot is dominating social media with its frighteningly good essays | CNN Business*. CNN. https://www.cnn.com/2022/12/05/tech/chatgpt-trnd/index.html

Kelly, S. M. (2023, January 26). *ChatGPT passes exams from law and business schools | CNN Business*. CNN. https://www.cnn.com/2023/01/26/tech/chatgpt-passes-exams/index.html

Kim, J., Bae, J., & Hastak, M. (2018). Emergency information diffusion on online social media during storm Cindy in U.S. *International Journal of Information Management*, *40*, 153–165. https://doi.org/10.1016/j.ijinfomgt.2018.02.003

Koetsier, J. (2023, March 14). *GPT-4 Beats 90% Of Lawyers Trying To Pass The Bar*. Forbes. https://www.forbes.com/sites/johnkoetsier/2023/03/14/gpt-4-beats-90-of-lawyers-trying-to-pass-the-bar/

Korn, J. (2023, February 22). *Vanderbilt University apologizes for using ChatGPT to write mass-shooting email | CNN Business*. CNN. https://www.cnn.com/2023/02/22/tech/vanderbilt-chatgpt-shooting-email/index.html

Lee, H. (2023). The rise of ChatGPT: Exploring its potential in medical education. *Anatomical Sciences Education*, ase.2270. https://doi.org/10.1002/ase.2270

Leiter, C., Zhang, R., Chen, Y., Belouadi, J., Larionov, D., Fresen, V., & Eger, S. (2023). *ChatGPT: A Meta-Analysis after 2.5 Months*. https://doi.org/10.48550/ARXIV.2302.13795

Li, L., Fan, L., Atreja, S., & Hemphill, L. (2023). *"HOT" ChatGPT: The promise of ChatGPT in detecting and discriminating hateful, offensive, and toxic comments on social media*. https://doi.org/10.48550/ARXIV.2304.10619

Li, L., Johnson, J., Aarhus, W., & Shah, D. (2022). Key factors in MOOC pedagogy based on NLP sentiment analysis of learner reviews: What makes a hit. *Computers & Education*, *176*, 104354. https://doi.org/10.1016/j.compedu.2021.104354

Li, L., Ma, Z., & Cao, T. (2021). Data-driven investigations of using social media to aid evacuations amid Western United States wildfire season. *Fire Safety Journal*, *126*, 103480. https://doi.org/10.1016/j.firesaf.2021.103480

Lim, W. M., Gunasekara, A., Pallant, J. L., Pallant, J. I., & Pechenkina, E. (2023). Generative AI and the future of education: Ragnarök or reformation? A paradoxical perspective from management educators. *The International Journal of Management Education*, *21*(2), 100790. https://doi.org/10.1016/j.ijme.2023.100790

Liu, Y., Ott, M., Goyal, N., Du, J., Joshi, M., Chen, D., Levy, O., Lewis, M., Zettlemoyer, L., & Stoyanov, V. (2019). *RoBERTa: A Robustly Optimized BERT Pretraining Approach*. https://doi.org/10.48550/ARXIV.1907.11692

Loureiro, D., Barbieri, F., Neves, L., Anke, L. E., & Camacho-Collados, J. (2022). *TimeLMs: Diachronic Language Models from Twitter*. https://doi.org/10.48550/ARXIV.2202.03829

Lund, B. D., Wang, T., Mannuru, N. R., Nie, B., Shimray, S., & Wang, Z. (2023). ChatGPT and a new academic reality: Artificial Intelligence-written research papers and the ethics of the large language models in scholarly publishing. *Journal of the Association for Information Science and Technology*, asi.24750. https://doi.org/10.1002/asi.24750

MacNeil, S., Tran, A., Mogil, D., Bernstein, S., Ross, E., & Huang, Z. (2022). Generating Diverse Code Explanations using the GPT-3 Large Language Model. *Proceedings of the 2022 ACM Conference on International Computing Education Research - Volume 2*, 37–39. https://doi.org/10.1145/3501709.3544280

Malik, A., Khan, M. L., & Hussain, K. (2023). How is ChatGPT Transforming Academia? Examining its Impact on Teaching, Research, Assessment, and Learning. *SSRN Electronic Journal*. https://doi.org/10.2139/ssrn.4413516

McInnes, L., Healy, J., & Melville, J. (2018). *UMAP: Uniform Manifold Approximation and Projection for Dimension Reduction*. https://doi.org/10.48550/ARXIV.1802.03426

Meckler, L., & Verma, P. (2022, December 29). Teachers are on alert for inevitable cheating after release of ChatGPT. *Washington Post*. https://www.washingtonpost.com/education/2022/12/28/chatbot-

cheating-ai-chatbotgpt-teachers/

Megahed, F. M., Chen, Y.-J., Ferris, J. A., Knoth, S., & Jones-Farmer, L. A. (2023). *How Generative AI models such as ChatGPT can be (Mis)Used in SPC Practice, Education, and Research? An Exploratory Study*. https://doi.org/10.48550/ARXIV.2302.10916

Mhlanga, D. (2023). Open AI in Education, the Responsible and Ethical Use of ChatGPT Towards Lifelong Learning. *SSRN Electronic Journal*. https://doi.org/10.2139/ssrn.4354422

Milmo, D. (2023, February 2). ChatGPT reaches 100 million users two months after launch. *The Guardian*. https://www.theguardian.com/technology/2023/feb/02/chatgpt-100-million-users-open-ai-fastest-growing-app

Mintz, M. P. R., Steven. (2023, January 15). *ChatGPT: Threat or Menace?* https://www.insidehighered.com/blogs/higher-ed-gamma/chatgpt-threat-or-menace

Mislove, A., Lehmann, S., Ahn, Y.-Y., Onnela, J.-P., & Rosenquist, J. (2021). Understanding the Demographics of Twitter Users. *Proceedings of the International AAAI Conference on Web and Social Media*, *5*(1), 554–557. https://doi.org/10.1609/icwsm.v5i1.14168

Moore, S., Nguyen, H. A., Bier, N., Domadia, T., & Stamper, J. (2022). Assessing the Quality of Student-Generated Short Answer Questions Using GPT-3. In I. Hilliger, P. J. Muñoz-Merino, T. De Laet, A. Ortega-Arranz, & T. Farrell (Eds.), *Educating for a New Future: Making Sense of Technology-Enhanced Learning Adoption* (Vol. 13450, pp. 243–257). Springer International Publishing. https://doi.org/10.1007/978-3-031-16290-9_18

Nguyen, D. Q., Vu, T., & Nguyen, A. T. (2020). *BERTweet: A pre-trained language model for English Tweets*. https://doi.org/10.48550/ARXIV.2005.10200

OpenAI. (2022a, November 2). *DALL·E API now available in public beta*. https://openai.com/blog/dall-e-api-now-available-in-public-beta

OpenAI. (2022b, November 30). *Introducing ChatGPT*. https://openai.com/blog/chatgpt

OpenAI. (2023, March 14). *GPT-4*. https://openai.com/research/gpt-4

Park, Y.-H., Choi, Y.-S., Park, C.-Y., & Lee, K.-J. (2022). EssayGAN: Essay Data Augmentation Based on Generative Adversarial Networks for Automated Essay Scoring. *Applied Sciences*, *12*(12), 5803. https://doi.org/10.3390/app12125803

Pavlik, J. V. (2023). Collaborating With ChatGPT: Considering the Implications of Generative Artificial Intelligence for Journalism and Media Education. *Journalism & Mass Communication Educator*, *78*(1), 84–93. https://doi.org/10.1177/10776958221149577

Pichai, S. (2023, February 6). *An important next step on our AI journey*. Google. https://blog.google/technology/ai/bard-google-ai-search-updates/

Powell, J. (2015). *A librarian's guide to graphs, data and the semantic web* (1st edition). Elsevier.

Qadir, J. (2022). *Engineering Education in the Era of ChatGPT: Promise and Pitfalls of Generative AI for Education* [Preprint]. https://doi.org/10.36227/techrxiv.21789434.v1

Reimers, N., & Gurevych, I. (2019). *Sentence-BERT: Sentence Embeddings using Siamese BERT-Networks*. https://doi.org/10.48550/ARXIV.1908.10084

Roose, K. (2023, January 12). Don't Ban ChatGPT in Schools. Teach With It. *The New York Times*. https://www.nytimes.com/2023/01/12/technology/chatgpt-schools-teachers.html

Rosenblatt, K. (2023, January 23). *ChatGPT passes MBA exam given by a Wharton professor*. NBC News. https://www.nbcnews.com/tech/tech-news/chatgpt-passes-mba-exam-wharton-professor-rcna67036

Rudolph, J., Tan, S., & Tan, S. (2023). ChatGPT: Bullshit spewer or the end of traditional assessments in higher education? *Journal of Applied Learning & Teaching*, *6*(1). https://doi.org/10.37074/jalt.2023.6.1.9

Ryan-Mosley, T. (2023, March 27). *An early guide to policymaking on generative AI*. MIT Technology Review. https://www.technologyreview.com/2023/03/27/1070285/early-guide-policymaking-generative-ai-gpt4/

Sallam, M. (2023). ChatGPT Utility in Healthcare Education, Research, and Practice: Systematic Review on the Promising Perspectives and Valid Concerns. *Healthcare*, *11*(6), 887. https://doi.org/10.3390/healthcare11060887

Sallam, M., Salim, N., Barakat, M., & Al-Tammemi, A. (2023). ChatGPT applications in medical, dental, pharmacy, and public health education: A descriptive study highlighting the advantages and limitations. *Narra J*, *3*(1), e103. https://doi.org/10.52225/narra.v3i1.103

Shepardson, D., & Bartz, D. (2023, April 12). *US begins study of possible rules to regulate AI like ChatGPT | Reuters*. Reuters. https://www.reuters.com/technology/us-begins-study-possible-rules-regulate-ai-like-chatgpt-2023-04-11/

Shravya Bhat, Nguyen, H., Moore, S., Stamper, J., Sakr, M., & Nyberg, E. (2022). *Towards Automated Generation and Evaluation of Questions in Educational Domains*. https://doi.org/10.5281/ZENODO.6853085

Smith, M., Ceni, A., Milic-Frayling, N., Shneiderman, B., Mendes Rodrigues, E., Leskovec, L., & Dunne, C. (2010). *NodeXL: a free and open network overview, discovery and exploration add-in for Excel 2007/2010/2013/2016 from the Social Media Research Foundation*. https://www.smrfoundation.org/nodexl/faq/how-do-i-cite-nodexl-in-my-research-publication/

Sok, S., & Heng, K. (2023). ChatGPT for Education and Research: A Review of Benefits and Risks. *SSRN Electronic Journal*. https://doi.org/10.2139/ssrn.4378735

Stokel-Walker, C. (2022). AI bot ChatGPT writes smart essays—Should professors worry? *Nature*. https://doi.org/10.1038/d41586-022-04397-7

Sun, G. H., & Hoelscher, S. H. (2023). The ChatGPT Storm and What Faculty Can Do. *Nurse Educator*, *48*(3), 119–124. https://doi.org/10.1097/NNE.0000000000001390

Tack, A., & Piech, C. (2022). *The AI Teacher Test: Measuring the Pedagogical Ability of Blender and GPT-3 in Educational Dialogues*. https://doi.org/10.5281/ZENODO.6853187

Taecharungroj, V. (2023). "What Can ChatGPT Do?" Analyzing Early Reactions to the Innovative AI Chatbot on Twitter. *Big Data and Cognitive Computing*, *7*(1), 35. https://doi.org/10.3390/bdcc7010035

Thorp, H. H. (2023). ChatGPT is fun, but not an author. *Science*, *379*(6630), 313–313. https://doi.org/10.1126/science.adg7879

Thurzo, A., Strunga, M., Urban, R., Surovková, J., & Afrashtehfar, K. I. (2023). Impact of Artificial Intelligence on Dental Education: A Review and Guide for Curriculum Update. *Education Sciences*, *13*(2), 150. https://doi.org/10.3390/educsci13020150

Tlili, A., Shehata, B., Adarkwah, M. A., Bozkurt, A., Hickey, D. T., Huang, R., & Agyemang, B. (2023). What if the devil is my guardian angel: ChatGPT as a case study of using chatbots in education. *Smart Learning Environments*, *10*(1), 15. https://doi.org/10.1186/s40561-023-00237-x

Tracy, R. (2023, April 11). *Biden Administration Weighs Possible Rules for AI Tools Like ChatGPT - WSJ*. The Wall Street Journal. https://www.wsj.com/articles/biden-administration-weighs-possible-rules-for-ai-tools-like-chatgpt-46f8257b

Twitter Inc. (2023). *About different types of Tweets*. https://help.twitter.com/en/using-twitter/types-of-tweets

Ventayen, R. J. M. (2023). OpenAI ChatGPT Generated Results: Similarity Index of Artificial Intelligence-Based Contents. *SSRN Electronic Journal*. https://doi.org/10.2139/ssrn.4332664

Zhai, X. (2022). ChatGPT User Experience: Implications for Education. *SSRN Electronic Journal*. https://doi.org/10.2139/ssrn.4312418

Zhu, M., Liu, O. L., & Lee, H.-S. (2020). The effect of automated feedback on revision behavior and learning gains in formative assessment of scientific argument writing. *Computers & Education*, *143*, 103668. https://doi.org/10.1016/j.compedu.2019.10366835